\documentclass[11pt]{article}

\usepackage{hyperref}       % hyperlinks
\usepackage{url}            % simple URL typesetting

\usepackage{nicefrac}       % compact symbols for 1/2, etc.
\usepackage{microtype}

\usepackage{bm}
\usepackage{mathtools}

\DeclarePairedDelimiter\floor{\lfloor}{\rfloor}

\usepackage{amsmath,amssymb,graphicx,amsopn,amsfonts}
\usepackage{algorithm}
\usepackage{algorithmic}
\usepackage{xcolor}
\usepackage{geometry}
\usepackage{tabularx}
\newcolumntype{b}{>{\hsize=1.5\hsize}X}
\newcolumntype{s}{>{\hsize=.45\hsize}X}

\newcommand*\Laplace{\mathop{}\!\mathbin\bigtriangleup}

\newtheorem{thm}{Theorem}[section]

\newcommand{\dt}{\Delta t}
\newcommand{\dw}{\Delta W}

\newcommand{\wh}{\widehat}

\newcommand{\eg}{{e.g.}}

\newcommand{\dsp}{\displaystyle}

\begin{document}
\title{TITLE}
\title{Random Batch Algorithms for Quantum Monte Carlo simulations}
\author{Shi Jin\footnote{School of Mathematical Sciences, Institute of Natural Sciences, MOE-LSEC and SHL-MAC, Shanghai Jiao Tong University, Shanghai, China (shijin-m@sjtu.edu.cn)}, Xiantao Li\footnote{Department of Mathematics, Pennsylvania State University, University Park, PA 16802, USA (Xiantao.Li@psu.edu)}}
\maketitle

\begin{abstract}
Random batch algorithms are constructed for quantum Monte Carlo simulations. The main objective is to alleviate the computational cost associated with the calculations of two-body interactions, including the pairwise interactions in the potential energy,  and the two-body terms in the Jastrow factor. In the framework of variational Monte Carlo methods, the random batch algorithm is constructed based on the over-damped Langevin dynamics, so that updating the position of each particle in an $N$-particle system only requires $\mathcal{O}(1)$ operations, thus for each time step the computational cost for $N$ particles is reduced from $\mathcal{O}(N^2)$ to $\mathcal{O}(N)$. For diffusion Monte Carlo methods, the random batch algorithm uses an energy decomposition to avoid the computation of the total energy in the branching step. The effectiveness of the random batch method is demonstrated using a  system of liquid ${}^4$He atoms interacting with a graphite surface.  
\end{abstract}

\section{Introduction}
One of the fundamental problems in chemistry is the computation of the ground state energy of a many-body quantum system. Although this major difficulty has been circumvented to some extent by the density-functional theory \cite{Kohn1965},  the quantum Monte 
Carlo (QMC) method \cite{anderson1975random,reynolds1982fixed,foulkes2001quantum,von1992quantum,anderson2007quantum} still remains an important approach to determine the ground state energy and electron correlations. 

This paper is concerned with the implementation of the QMC for many-body systems. More specifically, we consider the Hamiltonian,
\begin{equation}\label{eq: ham}
    \wh{H}= \sum_{i=1}^{N} -\frac{\hbar^2}{2m}{\triangle_{\bm r_i}} + \sum_{i\ne j} W(\bm r_i - \bm r_j) + \sum_{i=1}^{N} V_{ext}(\bm r_i).
\end{equation}
Here we use $\bm r=(\bm r_1, \bm r_2, \cdots, \bm r_N)$ to denote  the particle coordinates with $N$ being the total number of particles.  and the Laplacian ($-\Laplace$) in the first term of the Hamiltonian indicates the kinetic energy.  The second term in the Hamiltonian, which is a double sum, embodies the pairwise interactions, \eg, Coulomb, while the last term includes the external potential, namely,
\begin{equation}\label{eq: Vext}
    V_{ext}(\bm r_i) = \sum_{\alpha=1}^{M} U(\bm r_i - R_\alpha),
\end{equation}
where $R_\alpha$, for instance, can be the position of an atom. 

In principle, the ground state can be obtained by computing the smallest eigenvalue and the corresponding eigenfunction. 
It can be expressed in terms of a Rayleigh quotient, 
\begin{equation}\label{eq: E-var}
    E= \min_{\Phi} \frac{\dsp\int_{\mathbb{R}^{3N}} \Phi \wh{H} \Phi d\bm r_1 \cdots d \bm r_N} 
    {\dsp\int_{\mathbb{R}^{3N}} |\Phi|^2 d\bm r_1 \cdots d \bm r_N},
\end{equation}
and the minimizer $\Phi$ corresponds to the ground state  wave function.
However, due to the high dimensionality, a direct numerical approach, \eg, using finite difference or finite element methods together with numerical quadrature for the integrals suffers from the curse of dimensionality, thus is typically prohibitively expensive. 

Within the variational Monte Carlo (VMC) framework, this issue is addressed by selecting an appropriate ansatz, denoted here by  $\Phi \approx \Psi_0,$ for the many-body wave function. Then the multi-dimensional integral is interpreted as a statistical average, which can be sampled using a Monte Carlo procedure.
Traditionally, $\Psi_0$ is constructed using the one-body wave functions, with the effect of particle correlations described by Jastrow factors \cite{foulkes2001quantum}. 
Recently, artificial neural networks from machine learning have also been used to represent the many-body wave function \cite{carleo2017solving,han2020solving,han2019solving,pfau2019ab}. In fact, the recent surge of interest in applying  machine-learning algorithms to scientific computing problems has been a strong motivation for the current work.  

The first part of this paper is concerned with the numerical implementation of VMC. Since VMC formulates the energy calculation as a sampling problem, the most natural approach is the Metropolis-Hastings (MH) algorithm which, in general, falls into the category of Markov chain Monte Carlo (MCMC) algorithms in statistics. At each step, the chain is updated by calculating the energy change. As can be seen from \eqref{eq: ham} and \eqref{eq: E-var}, this requires visiting all  particles in the system. 	A direct treatment would involve $\mathcal{O}\big(N(N+M)\big)$ operations in each time step. 
The presence of the Jastrow factor further complicates the computation.  To alleviate the computational cost, we propose a random batch method (RBM), originated from emerging machine learning algorithms \cite{bottou1998online,wright2015coordinate,bubeck2014convex}, and recently introduced  to classical interacting particle systems in \cite{jin2020random} and extended to various applications in both
classical and quantum $N$-body systems \cite{GJP, jin2020mean,JLL2,KZ2020, li2020stochastic, li2020random, LLT}. In particular, \cite{jin2020random} established an error of RBM to be of  $O(\sqrt{\Delta t})$, where $\Delta t$ is the time step, {\it uniformly} in $N$. For the present problem, the objective is to use such an idea to quickly relax the quantum system and sample the energy in the VMC method. 

To this end, we first formulate the sampling problem using an over-damped Langevin equation, where the particles are driven by a drift and a stochastic force. The idea of using a Langevin dynamics to construct a VMC algorithm has been pursued in  \cite{scemama2006efficient}.  Rather than computing the particle interactions directly, our proposed RBM algorithm divides the system into random batches and only the interactions within each batch are computed. As a result, on average, updating all $N$ particles only requires $\mathcal{O}(N+M)$ operations. We justify the method by examining the transition density and show that at each step the density induced by the RBM is consistent with the exact transition kernel up to $\mathcal{O}(\dt^2)$, the same order as the Euler-Maruyama method.

The other important approach in QMC is the diffusion Monte Carlo (DMC) method  \cite{anderson1975random,reynolds1982fixed}, which starts with the time-dependent Schr\"odinger equation (TDSE), and evolves the quantum system in an imaginary time scale, leading to a parabolic equation \cite{reynolds1982fixed},
\begin{equation}\label{eq: tdse}
 \partial_t \Psi = (E_T-\wh{H} ) \Psi. 
\end{equation} 
The energy shift  $E_T$ is adjusted on-the-fly based on the change of the magnitude of the wave function. The key observation is that the dynamics \eqref{eq: tdse} can be associated with a stochastic process. In particular, the wave function $|\Psi|^2$ can be interpreted as the empirical measure of a particle system, in which the particles are driven by drift velocity and diffusion.  The growth/decay of the wave function is treated by introducing multiple copies of the system, each of which is called a walker or a diffuser  \cite{anderson1975random,reynolds1982fixed}. The number of walkers, which reflects the change of the norm of the wave function, is realized by using a birth/death process. The movement of the walkers is driven by the same over-damped Langevin dynamics. Therefore, the RBM is again a natural fit. On the other hand, the probability associated with the birth/death process depends on the total energy. To avoid the computation of the total energy $E$, especially before the ground state is reached, we propose to decompose the energy into one-, two-, and three-body terms. We construct an RBM where at each step a batch with three particles are selected and we only compute the energy within the batch. 

Speeding up QMC simulations has been an important focus in computational chemistry. Various software packages have been developed to this end \cite{scemama2012qmc,needs2020variational,kim2018qmcpack}. For instance, Kim et al. \cite{kim2018qmcpack} demonstrated how DMC algorithms can be efficiently implemented on high-performance computer clusters. They showed that when the dynamics of walkers is distributed among the OPENMP threads or MPI units, one can achieve an almost ideal speedup.  Toward this end,  we implemented the RBM algorithm by moving the  walkers in parallel, and we are able to perform QMC simulations of a Helium system with 5016 particles using only 60 cores.  

The rest of the paper is organized as follows. We first consider the RBM in the VMC setting in section 2, and justify the method in terms of the transition density.  Numerical results are presented for the Helium system. In section 3, we show the RBM in the DMC setting, followed by numerical results. The paper is concluded in section 4.

\section{The Random Batch Algorithm for the Variational Monte Carlo Methods}

The crucial observation that motivated the VMC framework is that the ground state energy can be viewed as an average with respective to a probability density,
\begin{equation}\label{eq: avgE}
    E = \big\langle  E_\text{tot}(\cdot) \big\rangle = \int p(\bm r) E_\text{tot}(\bm r) d\bm r,
\end{equation}
where $ p(\bm r)$ is regarded as a probability density function (PDF),
\begin{equation}\label{eq: pdf0}
    p(\bm r) \propto |\Phi_0(\bm r)|^2, 
\end{equation}
and the energy $E_\text{tot}$, given by,
\begin{equation}\label{eq: E0}
     E_\text{tot}(\bm r)= \frac{\wh{H} \Phi_0} {\Phi_0},
\end{equation}
will be regarded as a random variable.

The ground state wave function is usually sought in a Slater determinant form with a Jastrow factor \cite{jastrow1955many,foulkes2001quantum},
\begin{equation}\label{eq: ansatz0}
    \Phi_0=e^{-J(\bm r)} \Pi_{i=1}^{N} S\big(\phi(\bm r_1), \dots, \phi(\bm r_N)\big), \quad J(\bm r)= \sum_{i< j} u(|\bm r_i - \bm r_j|).
\end{equation}
Here $S$ is the Slater determinant with $\phi(\bm r)$ being the single-particle wave function, and we assume a common pairwise form $u(|\bm r_i - \bm r_j|$ for the Jastrow factor $J$.  It is also possible to include three-body terms. For simplicity, we do not consider the spin orbitals. 

We will consider Boson systems, which allow us to neglect the sign problem \cite{reynolds1982fixed} and focus exclusively on the sampling procedure. In addition, to have a class of explicit trial wave functions to work with, we follow the QMC methods for liquid Helium interacting with a graphite surface \cite{whitlock1998monte,pang2014diffusion}, where the following ansatz has been proven successful,
\begin{equation}\label{eq: ansatz}
    \Phi_0=e^{-J(\bm r)} \Pi_{i=1}^{N}  \phi(\bm r_i), \quad u(r)= \left( \frac{a}{r} \right)^5 + \frac{b^2}{r^2+c^2}.
\end{equation}
For homogeneous Hellium systems, the ansatz with only the Jastor factor has been widely used in QMC simulations \cite{kalos1974helium,mcmillan1965ground}. The ansatz in \eqref{eq: ansatz} includes orbitals centered around the graphite atoms.

From \eqref{eq: ansatz}, we can write the density \eqref{eq: pdf0} in an exponential form, 
\begin{equation}\label{eq: pr}
     p(\bm r) \propto e^{-2V}, \quad V= - \ln \Phi_0 = -  \sum_i \log \phi (\bm r_i) +  \frac12 \sum_{i} \sum_{j\ne i} u(|\bm r_i - \bm r_j|).
\end{equation}
The PDF is reminiscent of a Gibbs distribution with temperature $\beta^{-1}=1/2.$ 

The goal of VMC is to create samples according to such a probability density function, from which the ground state energy can be computed from \eqref{eq: avgE} by averaging over those samples.  Most VMC methods are of Markov chain Monte Carlo (MCMC) type. Namely, one constructs a Markov chain, which equilibrates to the PDF given by (or close to) \eqref{eq: pr}.  

Thanks to the explicit ansatz  \eqref{eq: ansatz} for the wave function, the total energy can be explicitly expressed as follows,
\begin{equation}\label{eq: E-tot}
E_\text{tot}(\bm r) =  \frac{1}{2} K_i^2 +   \sum_{i\ne j} W(\bm r_i - \bm r_j) + \sum_{i=1}^{N} V_{ext}(\bm r_i)
\end{equation}
Since the computational cost is of primary concern here, let us write out all the relevant terms. The first term comes from the kinetic energy, 
\begin{equation}\label{eq: Ki}
 K_i^2 = -\frac{\hbar^2}{m}  \frac{\triangle_i \Phi_0}{\Phi_0} = - \frac{\hbar^2}{m} \triangle_i \ln \Phi_0 -  \frac{\hbar^2}{m} | \nabla_i \ln \Phi_0  |^2. 
\end{equation} 
The actual form of the kinetic energy depends on the choice of the ansatz for $\Phi$. For instance, with the choice \eqref{eq: pr},the total energy is given by
\begin{equation}\label{eq: E-tot'}
E_\text{tot}(\bm r) =  - \frac{\hbar^2}{2m}    \triangle V -  \frac{\hbar^2}{2m}  \| \nabla V\|^2  +   \sum_{i\ne j} W(\bm r_i - \bm r_j) + \sum_{i=1}^{N}   \sum_{\alpha=1}^{M} U(\bm r_i - R_\alpha).
\end{equation}

Since the one-particle wave function is non-negative, we express it as exponential functions,
\begin{equation}\label{eq: phii}
\phi(\bm r_i) =  \sum_{\alpha=1}^M e^{-\theta(\bm r_i - R_\alpha)},
\end{equation} 
for some function $\theta$. This form has been used in \cite{whitlock1998monte} and the parameters were obtained by solving a one-dimensional Schr\"odinger equation.
%Here $M$ is the number of nuclei in the system. 

In light of \eqref{eq: E-tot'}, the calculation of the total energy, which will be part of both the variational and diffusion Monte Carlo algorithms,  scales {\it quadratically} in terms of the number of particles $N$.

\subsection{The classical Metropolis-Hastings Algorithm}

A classical algorithm in VMC is the Metropolis-Hastings algorithm. This algorithm is usually implemented by randomly displacing one particle as a time. With the observation that,
\begin{equation}\label{eq: Vi}
    V= \sum_i V_i, \quad V_i= -\log \phi (\bm r_i) +  \sum_{\overset{j=1}{j\ne i}}^N u(|\bm r_i - \bm r_j|), 
\end{equation}
only $V_i$ needs to be computed to determine the energy change due to the change of $\bm r_i$, which subsequently determines the rejections/acceptance of this move. The MH algorithm is standard in computational chemistry for both classical  and quantum systems \cite{allen2017computer}, so we keep the discussion brief and summarize the algorithm in {\bf Algorithm} \ref{alg:mh}.  Notice that the only parameters in the algorithm are the size of the trial moves, denoted by $\Delta x$, $\Delta y$ and $\Delta z$ in each of the three spatial directions, respectively.

\begin{algorithm}
\caption{Metropolis-Hastings (MH) algorithm for variational Monte Caro }
\label{alg:mh}
\begin{algorithmic}
\FOR{ nt=1, num\_steps}
\FOR{np=1, num\_particles}
\STATE{ Randomly pick an atom $i$ }
\STATE{ e\_old = $V_i$  in \eqref{eq: Vi}; }
\STATE{ $\bf r$\_old = $\bf r$\_i; }
\STATE{ ${\bf r}$\_i $\leftarrow$ ${\bf {r}}$\_i +  ( (rand() -0.5)*$\Delta$x, (rand() -0.5)*$\Delta
$y, (rand() -0.5)*$\Delta$z ); }
\STATE{ Compute the energy e\_new= $V_i$ and $\Delta$E = e\_new - e\_old; }
\IF{ exp[-$2\Delta$E] $>$ rand()   }
\STATE{$\bf r$\_i= $\bf r$\_old}
\ENDIF
\ENDFOR
\ENDFOR
\end{algorithmic}
\end{algorithm}

It is clear from  \eqref{eq: phii} and \eqref{eq: Vi}  that updating the position of one particle requires $\mathcal{O}(N+M)$ operations. Our goal is to reduce the cost of this computation to $\mathcal{O}(1).$

\subsection{A random batch algorithm based on the over-damped Langevin Dynamics}

The idea behind  the random batch algorithm can be best explained in terms of an over-damped Langevin dynamics, 
\begin{equation}\label{eq: lgv}
    d {\bm r}_i =  \nabla \log \phi(\bm r_i) dt -  \sum_{\overset{j=1}{j\ne i}}^N \nabla_{\bm r_i} u(|\bm r_i - \bm r_j|) dt +  dW_i(t), \quad 1 \le i \le N.
\end{equation}
 Here $W_i(t)$'s are independent Wiener processes. 
 Its empirical measure $f(\bm r,t)$ corresponds to the Fokker Planck equation (FPE),
\begin{equation}\label{eq: fpe}
    \partial_t f = - \nabla \cdot  
    \big( \bm v f \big) + \frac12 \triangle f,
\end{equation}
where $\bm v= (\bm v_1, \bm v_2, \ldots, \bm v_N)$ and
\begin{equation}\label{eq: vi}
\bm v_i=\nabla \log \phi(\bm r_i) -  \sum_{\overset{j=1}{j\ne i}}^N \nabla_{\bm r_i} u(|\bm r_i - \bm r_j|),\end{equation}
is interpreted as a drift velocity.
Under suitable conditions \cite{Mattingly:02}, the dynamical system with potential given by \eqref{eq: pr} is ergodic, and the PDF $p(\bm r)$ in \eqref{eq: pr} is the unique equilibrium measure of this stochastic system. Therefore the numerical integration of the SDEs \eqref{eq: lgv} offers a route to navigate to  \eqref{eq: pr} and sample the energy.  

Using the over-dampled Langevin equation to sample the Gibb distribution has been a widely known method. In the context of VMC,  this approach has been adopted by Scemama et al. \cite{scemama2006efficient} to improve standard methods. In addition, they combined the Langevin dynamics with the Metropolis-Hastings algorithm to accept/reject the produced samples.   

A direct discretization, e.g., the Euler-Maruyama method \cite{kloeden2013numerical}, would involve the following step \cite{kloeden2013numerical},
\begin{equation}\label{eq: EM}
   {\bm r}_i(t+\dt) = {\bm r}_i(t) +    \nabla \log \phi(\bm r_i) \dt -  \sum_{j\ne i} \nabla_{\bm r_i} u(|\bm r_i(t) - \bm r_j(t)|) \dt + \dw_i,  \quad 1 \le i \le N.
\end{equation}
Here we assume that the step size $\dt$ is uniform, and the discrete time  is given by $\mathcal{T} := \{n\dt, n\geq 0 \}.$ The method \eqref{eq: EM} is applied to each time step $t\in \mathcal{T}$. At each step, $\dw_i$ is sampled from a normal random distribution with zero mean and variance $\dt.$

Although the Euler-Maruyama method is completely different from the Metropolis-Hastings algorithm, they nevertheless have a similar  computational cost for updating the position of each particle.  More specifically, one has to compute the interactions with all other particles ($ u(|\bm r_i(t) - \bm r_j(t)|) $), for all $j \neq i$. In addition, one needs to compute $\log \phi(\bm r_i)$, which is given by,
\begin{equation}\label{eq: log-phi}
\log \phi(\bm r_i) = \log \sum_{\alpha=1}^M e^{-\theta(\bm r_i - R_\alpha)}. 
\end{equation}
Together, they contribute to  $\mathcal{O}(M+N)$ operations for each particle at each time step. 

\medskip

To reduce the cost of evaluating the two-body interactions, the RBM proceeds as follows (this corresponds to the RBM with replacement in \cite{jin2020random}): At each step, one randomly picks out two particles, $i$ and $j$, and compute their interactions,
 $\nabla_{\bm r_i} u(|\bm r_i - \bm r_j|)$,
 then updates their positions as follows, 
\begin{equation}
\label{eq-1}
\left\{
\begin{aligned}
    {\bm r}_i(t+\dt) &= {\bm r}_i(t) + \nabla \log \phi(\bm r_i) \dt  + (N-1) \nabla_{\bm r_i} u(|\bm r_i - \bm r_j|)  \dt +  \dw_i, \\
    {\bm r}_j(t+\dt) &= {\bm r}_j(t) + \nabla \log \phi(\bm r_j) \dt  + (N-1) \nabla_{\bm r_j} u(|\bm r_i - \bm r_j|)  \dt +  \dw_j.
\end{aligned}
\right.
\end{equation}
Notice that $\nabla_{\bm r_j} u(|\bm r_i - \bm r_j|)=-\nabla_{\bm r_i} u(|\bm r_i - \bm r_j|) $, thus only one of them needs to be computed.
The factor $(N-1)$ accounts for the fact that we are using {\it one} term $u(|\bm r_i - \bm r_j|)$ to account for the interactions with all $(N-1)$ particles. In general, it is also possible to pick larger random batches. Choosing batches with two particles is most popular.  

In light of \eqref{eq: log-phi}, the computation of the one-body term still involves $\mathcal{O}(M)$ operations. However, since
\begin{equation}\label{eq: 1-body}
   \nabla \log \phi(\bm r_i)  = \dsp \sum_{\alpha=1}^M - \nabla \theta(\bm r_i - R_\alpha) q_\alpha^i, \quad q_\alpha^i = \frac{e^{-\theta(\bm r_i - R_\alpha)}}{\sum_{\beta=1}^M e^{-\theta(\bm r_i - R_\beta)}},
\end{equation}
where the coefficients $q_\alpha^i$'s are non-negative and $\sum_\alpha q_\alpha^i =1,$ thus the log-gradient term can be viewed as a statistical average with discrete probability given by $\left\{q_\alpha^i\right\}_{\alpha=1}^M.$ So a simple idea is to pick {\it just one} term $\alpha$ randomly, \eg, by using a direct Monte  Carlo method for one step. The implementation is straightforward: Assume that one starts with $\alpha$ and  computes
$e_{old}=\theta(\bm r_i - R_\alpha)$, and then we randomly pick $1 \le \beta \le M$, and compute $e_{new}=\theta(\bm r_i - R_\beta)$. We accept $\beta$ with probability $\exp\big[e_{new}-e_{old}\big].$ 
\medskip
We summarize the random batch algorithm in {\bf Algorithm 2}.
\begin{algorithm}
\caption{Random batch algorithm for variational Monte Carlo }
\label{alg:rb-vmc}
\begin{algorithmic}
\FOR{nt=1, num\_steps}

\FOR{np=1, num\_particles/2}
    \STATE{Randomly pick two particles $i$ and $j$ with $i\neq j$.}
     \STATE{Perform one step of the Monte Carlo algorithm with respect to  $\left\{q_\alpha^i\right\}$ and select $\alpha$. Compute $\bm b_i=- \nabla \theta(\bm r_i - R_\alpha)$.}  
          \STATE{Perform one step of the Monte Carlo algorithm with respect to  $\left\{q_\alpha^j\right\}$ and select $\beta$. Compute $\bm b_j=- \nabla \theta(\bm r_j - R_\beta)$.}  
    \STATE{Evaluate $\bm u_{ij}= -\bm u_{ji}=  (N-1) \nabla_{\bm r_i} u(|\bm r_i - \bm r_j|).$}
    \STATE{Update the particle positions,
    \begin{equation}
 \begin{aligned}
     {\bm r}_i \longleftarrow &{\bm r}_i +  \bm b_i \dt + \bm u_{ij} \dt +  \dw_i, \\
     {\bm r}_j \longleftarrow &{\bm r}_j + \bm b_j \dt + \bm u_{ji} \dt +  \dw_j.
    \end{aligned}
\end{equation}}
\ENDFOR
\ENDFOR
\end{algorithmic}
\end{algorithm}

As a result of the random sampling of the one- and two-body interactions, updating the position of each particle {\it only requires $\mathcal{O}(1)$ operations}. In the next section, we will study the transition density of the random algorithm, which in turn serves as a validation of the algorithms.

\medskip
Another practical issue emerges when the interaction $u(|\bm r|) $ has a singularity near zero. In this case, a direct implementation of the random batch algorithm would often require much smaller step sizes in the integration of the Langevin dynamics \eqref{eq: lgv} \cite{li2020random}. The issue can be mitigated by separating $u(|\bm r|)$ into a singular, but short-ranged part, and a long-ranged, but smooth part \cite{li2020random}. The short-range interactions can be efficiently computed using Verlet's cell list method which, for each particle, still involves $\mathcal{O}(1)$ operations.  This is a common practice in classical molecular simulations \cite{allen2017computer,frenkel2001understanding}. Meanwhile, the long-range part,  which is where most computations are involved, can be simulated  by the random batch algorithm. Here we use a simple approach to separate out the singularity by introducing a cut-off distance $r_\text{cut}$, then replacing the short-range part by an extrapolation using a Taylor expansion, namely,
\begin{equation}\label{eq: uLS}
u_L(r) = \left\{
\begin{array}{ll}
u(r) & r > r_\text{cut}, \\
u(r_\text{cut}) + u'(r_\text{cut})(r-r_\text{cut}) + \frac12 u''(r_\text{cut})(r-r_\text{cut})^2, & \text{Otherwise}.
\end{array}\right.
\end{equation}

\begin{figure}[H]
\begin{center}
\includegraphics[scale=0.4]{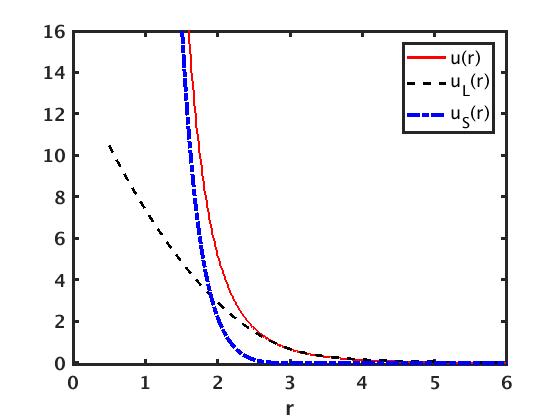}
\caption{Separation of the interaction $u(r)=r^{-5}$ with singularity at $r=0$ (solid line) into a long range interaction $u_L(r)$ (dashed) without singularity, and a short range interaction $u_S(r)$ (dot-dashed).}
\label{fig: ur}
\end{center}
\end{figure}
The short-range part is then defined as $u_S(r) = u(r) - u_L(r).$ Figure \ref{fig: ur} shows an example of how such a decomposition can be easily constructed.

%\bigskip
%
%Let us understand the random algorithm from the perspective of an MCMC method. In particular, we examine the transition kernels. From \eqref{eq: lgv}, we see that the transition kernel within one time step is given by,
%\begin{equation}
% p( \bm r(t+\dt) | \bm r (t) ) \propto  \Pi_{i=1}^N \exp - \frac{ \big(r_i(t+\dt) - r_i(t) - \nabla \log \phi(\bm r_i) \dt - \sum_{j\ne i} f_{ij} \dt\big)^2}{2 \sigma^2 \dt}.
%\end{equation}
%
%One the other hand, the transition kernel from the random batch algorithm is given by,
%\begin{equation}
%\begin{aligned}
%& p( \bm r(t+\dt) | \bm r (t) ) \propto
%   \Pi_{i,j=1, i\neq j }^N \\ & \exp - \frac{ \big(r_i(t+\dt) - r_i(t) - \nabla \log \phi(\bm r_i) \dt - (N-1) f_{ij} \dt\big)^2}{2 \sigma^2 \dt}\\
% & \times 
% \exp - \frac{ \big(r_j(t+\dt) - r_j(t) - \nabla \log \phi(\bm r_j) \dt - (N-1) f_{ji} \dt\big)^2}{2 \sigma^2 \dt},  \\
% \approx & \Pi_{i=1}^N  \exp - \frac{ \big(r_i(t+\dt) - r_i(t) - \nabla \log \phi(\bm r_i) \dt - \sum_{j\ne i} f_{ij} \dt\big)^2}{2 (\sigma/(N-1))^2 \dt}\
% \end{aligned}
%\end{equation}

\subsection{The transition kernel  of the random batch algorithm}

\subsubsection{The random batch algorithm for the one-body term}

We will first consider the Monte-Carlo sampling of the one-body term \eqref{eq: 1-body}, and for clarity we place the problem  in the setting of solving a $d$-dimensional SDE system,
\begin{equation}
 d\bm r(t)= \bm a(\bm r(t)) dt + \sigma dW_t.
\end{equation}
Here $\sigma \geq 0$ is a constant, which is also allowed to be zero. In light of \eqref{eq: 1-body}, we consider a vector field $\bm a$ 
that can be expressed as,
\begin{equation}
 \bm a(\bm r) = \sum_{\alpha=1}^M q_\alpha \bm a_\alpha(\bm r),
\end{equation}
where the coefficients $q_\alpha$'s represent a discrete probability density, that is, $q_\alpha \geq 0$ and $\sum_\alpha q_\alpha =1.$ We examine the random algorithm,
\begin{equation}\label{eq: rand-1}
\bm r(t+\dt) = \bm r(t) + \bm a_\alpha\big(\bm r(t) \big) \dt + \sigma \dw,
\end{equation}
where the index $\alpha$ is selected at random according to the discrete density. We consider uniform step size $\dt$, and the equation will be applied to each step $t.$

Clearly, the corresponding transition density is given by,
\begin{equation}
 p\big( \bm r(t+\dt)=\bm y| \bm r(t) =\bm x\big) =\sum_{\alpha=1}^M q_\alpha \frac{1}{(2\pi\sigma^2)^{d/2}} \exp \left[-\frac{\big(\bm y-  \bm x - \bm a_\alpha(\bm x) \dt \big)^2}{2\sigma^2}
 \right]. 
\end{equation}

For any function $A(\bm r) \in C^4(\mathbb{R}^d)$ with suitable growth conditions \cite{kloeden2013numerical}, one has,
\begin{equation}
\begin{aligned}
&\int_{\mathbb{R}^d}  A(\bm y)  p\big( \bm x(t+\dt)=\bm y| \bm x(t) =\bm x\big) d\bm y \\
& =\sum_{\alpha=1}^M q_\alpha \big[ A(\bm x) + \bm a_\alpha(\bm x)  \cdot \nabla A(\bm x) \dt 
 + \frac12 \triangle A(x) \dt + \mathcal{O}(\dt^2) \big]\\
 =& A(\bm x) + \bm a(\bm x)  \cdot \nabla A(\bm x) \dt 
 + \frac12 \triangle A(x) \dt + \mathcal{O}(\dt^2).
\end{aligned}
\end{equation}
Therefore, this random algorithm has a first weak-order of accuracy, which is  comparable  to the Euler-Maruyama method. Even though the drift term $\bm a(\bm r)$ is only sampled once at each step, the method is still convergent. To our knownledge, this surprising property was first noticed by E et al. in the context of multiscale methods for SDEs \cite{weinan2005analysis}, where the weak convergence is proved in a more general (multiscale) setting.

\subsubsection{The random batch algorithm for pair-wise interactions}
We now turn to the SDE system \eqref{eq: lgv} with pair-wise interactions, 
\begin{equation}\label{eq: sde-pair}
  d\bm r_i(t) = \nabla \log \phi (\bm r_i) dt -   \sum_{j \neq i} \nabla u (|\bm r_i - \bm r_j|) dt + dW_t.
\end{equation}

By letting $\bm u_{ij}= \nabla u (|\bm r_i - \bm r_j|)$, we can write the pair-wise terms as, 
\begin{equation}
\bm u_i = \sum_{j \neq i} \bm u_{ij}, \;\; \bm u_{ij}=-\bm u_{ji}.
\end{equation}

To study the weak convergence, one may consider the conditional expectation,
\begin{equation}
 \mathbb{E} \Big[ A(\bm r(t+\dt) | \bm r(t)=\bm x \Big]. 
\end{equation}
This is represented by the transition density as follows,
\begin{equation}
\mathbb{E} \big[ A(\bm r(t+\dt) | \bm r(t)=\bm x \big] =\int A(\bm y) p\big(\bm r(t+\dt)=\bm y| \bm r(t)=\bm x \big) d\bm y. 
\end{equation}

The transition density for the SDEs \eqref{eq: sde-pair} follows the Fokker-Planck equation \cite{kloeden2013numerical}. The explicit form of the solution is often unknown. But with the approximation by the Euler-Maruyama method, 
\begin{equation}
 \bm r_i(t+\dt)= \bm r_i(t)  + \nabla \log \phi(\bm r_i) \dt  + \bm u_i \dt +  \Delta W_i,
\end{equation}
we can identify an approximate transition kernel,
\begin{equation}
\begin{aligned}
&p^{EM}(\bm r(t+\dt)=\bm y| \bm r(t)=\bm x ) \\
&= \frac{1}{ ({2\pi \sigma^2\dt) }^{d/2}} \exp \left[ - \big(\bm y - \bm x -  \nabla \log \phi(\bm x) \dt - \bm u(\bm x) \dt \big)^2/(2\sigma^2 \dt) \right].  
\end{aligned}
\end{equation}

By the weak It\^o-Taylor expansion \cite{kloeden2013numerical}, we have from the density induced by the Euler-Maruyama method,
\begin{equation}\label{eq: pem}
 \mathbb{E} \big[ A(X(t+\dt) | X(t)=x \big] = A(x) + \mathcal{L}A(x) \dt + \mathcal{O}(\dt^2),
\end{equation}
where $\mathcal{L}$ is the generator, 
\begin{equation}
 \mathcal{L}A(\bm x) = \sum_i \big(\nabla \log \phi (\bm  x_i) + \bm  u_i \big) \cdot \nabla_{x_i} A(\bm  x)  + \frac12 \triangle A(\bm x). 
\end{equation}

The expansion \eqref{eq: pem} is consistent with that of the exact transition density up to $\mathcal{O}(\dt^2)$, making the Euler-Maruyama method
first order in the weak sense \cite{kloeden2013numerical}.

\bigskip

We now turn to the random batch algorithm \ref{eq-1} with replacement \cite{jin2020random}. The convergence property has recently been proved in \cite{JLL2}:
\begin{thm}
 The random batch algorithm over $N/2$ steps has weak order 1. 
\end{thm}

Here we illustrate the weak convergence in terms of the transition density. This also helps us to construct RBM for diffusion Monte Carlo. 
Since we randomly pick a pair of components to update, the transition density, denoted here by $p^{RB}$, is given by,
\begin{equation}\label{eq: prb}
  p^{RB}\big(\bm r(t+\dt)=\bm y| \bm r=\bm x \big) 
  = \frac{2}{(N-1)N} \sum_{i > j} q_{ij}(\bm y| \bm x),
\end{equation}
where,
\begin{equation}
\begin{aligned}
q_{ij}(\bm y| \bm x)= &\frac{1}{ ( 2\pi \dt)^{3N/2} } \exp \left[- \big(\bm y_i - \bm x_i - \nabla \log \phi(\bm x_i) \dt - (N-1) \bm u_{ij}) \dt \big)^2/(2 \dt) \right] \\
&\qquad \quad \times  \exp  \left[- \big(\bm y_j -\bm  x_j -\nabla \log \phi(\bm x_j) \dt - (N-1) \bm u_{ji} \dt \big)^2/(2 \dt) \right]\\
& \dsp \qquad \quad \times  \dsp\Pi_{k\ne i, j} \delta(\bm y_k-\bm x_k).
\end{aligned}
\end{equation}
The delta functions were included to ensure that when the pair $(i,j)$ is selected, other components are not updated. 
In the following discussions, we will simply write the transition density as $p^{RB}(\bm y|\bm x).$

With direct Taylor expansions, one finds that, for any observable $A(\bm x)$,
\begin{equation}
\begin{aligned}
 \int A(\bm y) q_{ij}(\bm y|\bm x) dy =& A(\bm x) + \nabla \log \phi(\bm x_i) \cdot \nabla_{\bm x_i} A(\bm x) \dt +\nabla \log \phi(\bm x_j) \cdot \nabla_{\bm x_j} A(\bm x) \dt   \\
    &+(N-1) \bm u_{ij} \cdot \nabla_{\bm x_i} A(\bm x) \dt +   (N-1) \bm u_{ji} \cdot \nabla_{\bm x_j} A(x) \dt  \\
    &+ \frac12  \triangle_{\bm x_i} A(\bm x) \dt + \frac12  \triangle_{\bm x_j} A(\bm x) \dt  +\mathcal{O} (\dt^2). 
\end{aligned}
\end{equation}

Combining this with \eqref{eq: prb},  we have,
\begin{equation}
\begin{aligned}
\int A(\bm y) p^{RB}(\bm y|\bm x) d\bm y =& A(\bm x)  + \frac{2\dt}{N}  \Big\{ \sum_i \nabla \log \phi(\bm x_i)\cdot  \nabla_{\bm x_i} A(\bm x)  \\
& +  \sum_i \sum_{j\ne i} \bm u_{ij} \cdot \nabla_{\bm x_i} A(\bm x) 
 + \frac12   \triangle A(\bm  x)\Big\} + \mathcal{O}(\dt^2).
\end{aligned}
\end{equation}

Therefore, the random batch algorithm with replacement, when applied to one batch of two particles, has the same accuracy as the Euler-Maruyama method over a time step of 
$2\dt/N.$ Note one full time step in Euler-Maruyama method corresponds to $N/2$ such steps in the RBM with replacement.

\subsection{Numerical Results}
\begin{figure}[htbp]
\centering
    \includegraphics[scale=0.4]{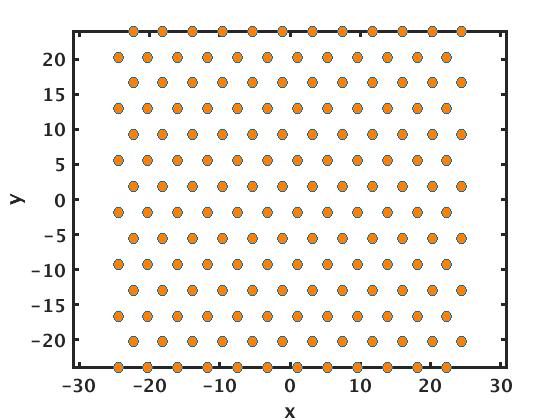}
    \caption{A two dimensional lattice with Helium atoms.}
    \label{fig: latt}
\end{figure}

We conduct numerical experiments with ${}^4$He atoms interacting with a two-dimensional lattice. The ${}^4$He atoms, due to the fact that the total spin is zero, are bosons. Driven by its superfluid properties and many observed quantum effects, ${}^4$He atoms have been extensively studied by computer simulations.  Acting as a substrate, the lattice  has a triangular structure with lattice spacing  given by $a_0= 4.2576 $ \AA.  Such a lattice can be generated using rectangular unit cells, each of which contains two atoms.  For example,  Figure \ref{fig: latt} shows such a system with $12\times 7$ unit cells and a total of 168 atoms.  The model is adapted from \cite{joly1992helium}. We choose  \AA~ as the length unit and $k_B $Kelvin as the unit of energy.  

Particles that represent the  wave function $\Phi_0$ are created randomly near the nuclei. We follow the setup in \cite{pang2014diffusion}. 
In particular, in the wave function ansatz \eqref{eq: ansatz}, the one-particle wave function is assumed to be,
\begin{equation}
 \phi(\bm r_i) = \exp -\big( (z_i - z_e)^2/z_0^2 \big) \sum_{\alpha=1}^M \exp \big( -(\bm r_i - R_\alpha)^2/r_0^2\big).
\end{equation} 
Here $z_i$ indicates the third component of the coordinate $\bm r_i.$ In addition, the two-body terms in the Jastrow factor are chosen to consist of both short and long range terms,
\begin{equation}\label{eq: ur}
 u(r) = \left( \frac{a}{r}\right)^5 + \frac{b^2}{c^2+r^2}.
\end{equation}
Although the first term decays rather quickly, we do not use an abrupt truncation of the function. Instead, we follow the construction \eqref{eq: uLS}, and split it into a function that vanishes beyond a cut-off distance $r_{cut}$. The remaining part is merged into the second term   in \eqref{eq: ur} and regarded as a long-range interaction. The parameters, with unit \AA, are given in Table \ref{params}. 

\begin{table}[thp]

\caption{ Model parameters in the QMC simulations of ${}^4$He. \label{params} }
\begin{tabularx}{\textwidth}{s|s|s|s|s|s|s}
% \begin{ruleredtabular}{@{\hspace{1em}}  c@{\hspace{1em}}   ccc @{\hspace{3em}} }
\hline\hline
 $z_e$ & $z_0$  & $r_0$ &  $a$  & b & c &$r_{cut}$    \\
 \hline
 2.85 &  0.521   &15 & 2.771 &   5.0 & 10.0   & 8.0\\
  \hline\hline
 \end{tabularx}
 \end{table}

We first carry out VMC simulations using RBM-VMC ({\bf Algorithm} \ref{alg:rb-vmc} ) and the Euler-Maruyama method \eqref{eq: EM}. In the simulations,  we run the algorithms with 300 ensembles and the average energy at each step will be computed as an average over these ensembles. In principle, the algorithms can be implemented with just one realization, and the ground state energy would be computed entirely from the time series. But multiple ensembles can be easily implemented in parallel. In addition, the ensembles can later be turned into walkers in the DMC simulations.

\begin{figure}[htbp]
\centering
    \includegraphics[scale=0.36]{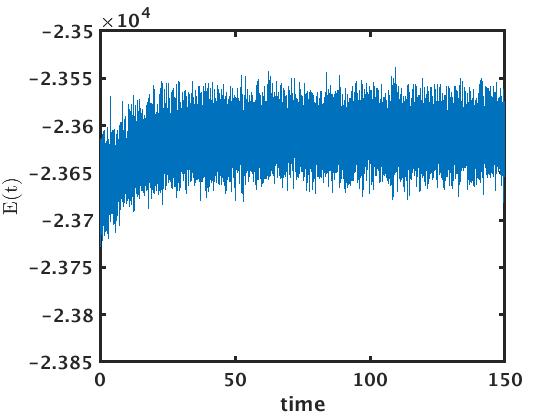}\\
       \includegraphics[scale=0.36]{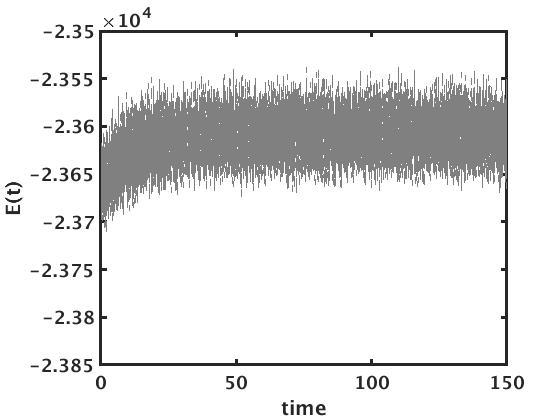}\\
    \includegraphics[scale=0.36]{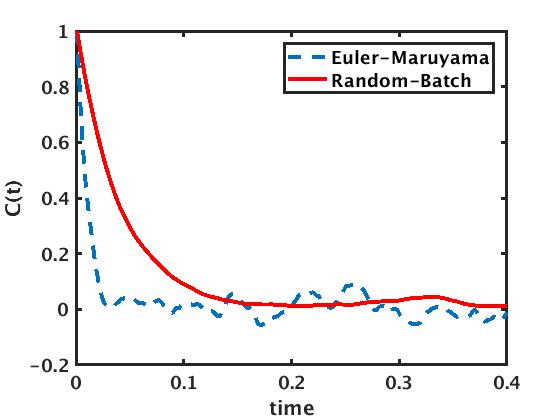}
    \caption{A comparison of the random batch Algorithm  \ref{alg:rb-vmc} (top) to the Euler-Maruyama method (middle). The bottom panel shows the time correlation. }
    \label{fig:vmc}
\end{figure}

 Figure \ref{fig:vmc} shows the average energy computed from the RBM-VMC and the Euler-Maruyama methods in the time interval [0,150]. The step size is $\dt= 10^{-3}.$   We observe that both methods relax to equilibrium around $t=25.$ Since the time scale is fictitious, we do not assign a unit for the time variable.

\begin{figure}[thp]
\centering
    \includegraphics[scale=0.42]{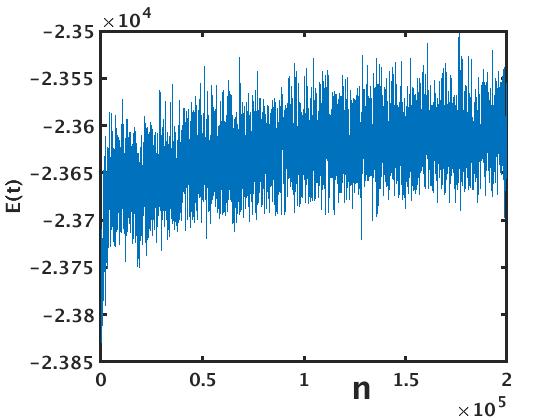}
    \caption{The energy sampled from 200,000 steps of the Metropolis-Hastings algorithm. }
    \label{fig:vmc-mh}
\end{figure}

We also show the time correlation of the sampled energy after the system has reached equilibrium.  To obtain a more quantitative comparison, we implemented an MCMC diagnostics. In this context, the relaxation is known as the burn-in period, and a thinning parameter can be used to indicate correlations. More specifically,  we use the Raftery and Lewis criteria \cite{raftery1992practical} ($q=0.025, r=0.0125,s=0.95$) and find that the burn-in period is 23.49 and 38.54, with thinning parameters 0.058 and 0.066, for the Euler-Maruyama and RBM, respectively. One can see that the random batch method has slightly longer burn-in time, and longer correlation. Since both of these methods are constructed by integrating SDEs in time, we have factored in the step size $\dt$ in estimating these parameters. We also show the energy sampled from the Metropolis-Hastings algorithm in Figure \ref{fig:vmc-mh}. The average energy is $2.361113 \times 10^4$ with standard statistical error $1.698.$ Note that it is not straightforward to compare the previous two algorithms to the Metropolis-Hastings algorithm, since the latter method does not have an associated time scale. 

We now compare the CPU time that is needed to move the 300 Markov chains for 1000 steps.  In this comparison, we have excluded the cost associated with the energy calculations in the random batch and Euler-Maruyama methods, since they are not needed in the burn-in period, and even upon equilibrium, it is a good practice to sample it every few steps to obtain less correlated samples. From Table \ref{vmc}, one clearly sees that the RBM is more efficient than the Euler-Maruyama method, mainly due to the random sampling of the pairwise interactions in the Jastrow factor in the wave function \eqref{eq: ansatz}. It is much more efficient than the Metropolis-Hastings algorithm, mainly because the latter method requires the calculation of the energy at {\it every } step. 
\begin{table}[thp]

\caption{ Comparison of the CPU time (measured in seconds) for several VMC methods. \label{vmc} }
\begin{tabularx}{\textwidth}{b|s|s|s}
% \begin{ruleredtabular}{@{\hspace{1em}}  c@{\hspace{1em}}   ccc @{\hspace{3em}} }
\hline\hline
  &  Metropolis-Hastings  & Euler-Maruyama &   Random Batch   \\
 \hline
 CPU time for a 1000-step sampling period  &  1503  & 469 &  54  \\
%\hline
 % CPU time without energy sampling  & 16478 & 1.0061 & 1094 \\
  \hline\hline
 \end{tabularx}
 \end{table}

\begin{figure}[htbp]
\centering
    \includegraphics[scale=0.2]{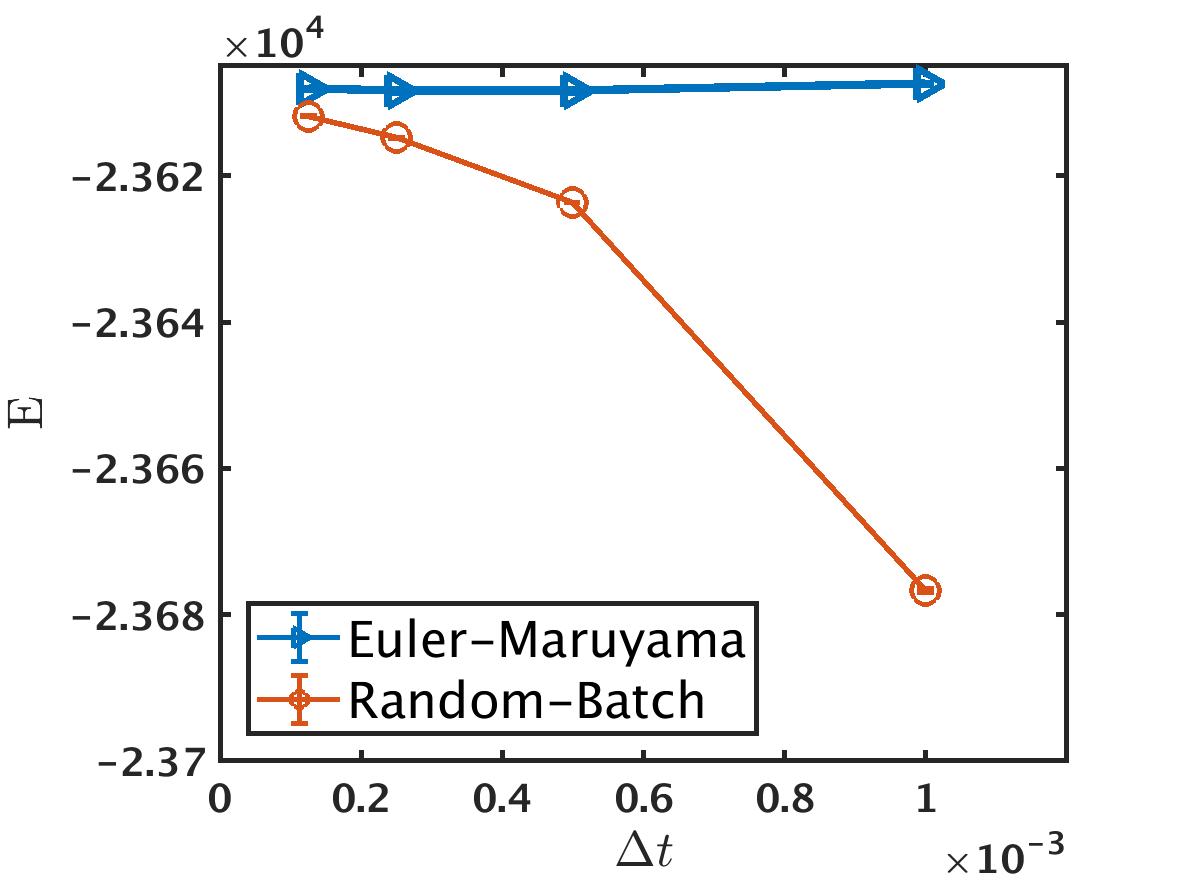} 
    \caption{The average of the energy computed from the random batch and Euler-Maruyama methods for various choices of the step size $\dt$. }
    \label{fig: ebar}
\end{figure}

Finally, we examine the effect of the time discretization. Unlike the metropolis-Hastings algorithm, the RBM and Euler-Maruyama methods are biased, and the results depend on the step size. Figure \ref{fig: ebar} shows the averages computed from the two methods for different choices of $\dt.$ We choose $10^5$ samples from equilibrium in the estimation. 
Compared to the values from the MH algorithm, it can be observed that the Euler-Maruyama method over-estimates the ground state energy, while the random  batch method under-estimates it.  

\section{The Random Batch Algorithm in Diffusion Quantum Monte Carlo Methods}

The accuracy of the VMC method is limited by the ansatz of the wave function \eqref{eq: ansatz}. The idea of the DMC is to go back to the time-dependent Schr\"{o}dinger equation and evolve the system along the imaginary time,
\begin{equation}\label{eq: imag-time}
    \partial_t \Psi = ( E_T - \widehat{H}) \Psi. 
\end{equation}
Here a rescaling of time scale $ i t/\hbar \to t$ has been introduced and $t$ now represents a fictitious time scale. Since the transient is not of interest here, we will not keep track of the time scales. 

Depending on the choice of the reference energy $E_T$, the solution would either decay or grow exponentially, unless $E_T$ coincides with the ground state energy, at which point, the wave function converges to the ground state as $t \to +\infty$. 

Instead of solving \eqref{eq: imag-time} directly, it is often more practical to find $f(\bm r,t)$ with 
\begin{equation}
    f(\bm r,t) = \Psi(\bm r,t) \Phi_0(\bm r).
\end{equation}
This ansatz has the flavor of the importance sampling. In addition, if one chooses  $\Psi(\bm r,0) = \Phi_0(\bm r)$, then  $f(\bm r, 0)= |\Phi_0|^2 \propto p(\bm r)$ in \eqref{eq: pr}. Therefore, we can use a VMC method to initialize $f(\bm r, t).$ 

Direct calculations yield the following differential equation \cite{reynolds1982fixed},
\begin{equation}\label{eq: f}
 \partial_t f = -\nabla\cdot \big( \frac{\hbar^2}{{m}} \bm v(\bm r) f\big) + \frac{\hbar^2}{2m} \nabla^2 f - \big(E_T - {E}_\text{tot}(\bm r) \big)  f.
\end{equation}

The average energy ${E}(t)$ is defined as a weighted average,
\begin{equation}
    {E}(t)= \frac{\dsp\int f(\bm r,t) E_\text{tot}(\bm r) d\bm r}{\dsp \int f(\bm r, t) d\bm r}.  
\end{equation}

Without the last term on the right hand side of (\ref{eq: f}), the equation above, with a time rescaling \(\tau \to \tau \hbar^2/m\),  would be reduced to the Fokker-Planck equation \eqref{eq: fpe}  associated 
with the SDE \eqref{eq: lgv}, with the additional term that embodies the influence of the choice of the energy shift on the change of total mass. 

Within a short time step, $\Delta t$, the solution of \eqref{eq: f} can be approximated by \cite{reynolds1982fixed},
\begin{equation}\label{eq: f'}
    f(\bm r,t+\Delta t) = \int_{\mathbb{R}^{3N} } G(\bm r, \bm r', \Delta t) 
    f(\bm r',t) d\bm r',
\end{equation}
where the function $G$, often referred to as  Green's function, is given by \cite{reynolds1982fixed},
\begin{equation}
     G(\bm r, \bm r', \Delta t) = \frac{1}{ \big( 2\pi \sigma^2 \big)^{3N/2}}
     \exp \left[ -  \frac{ \big(\bm r' - \bm r -  \frac{\Delta t \hbar}{m}  \bm v(\bm r)\big)^2 }{2\sigma^2} \right]  \exp 
     \left[\Delta t\big(E_T -  {E}_\text{tot}(\bm r)   \big)\right].  
\end{equation}
The parameter $\sigma =  \sqrt{\Delta t} \hbar/\sqrt{m}$ and the vector field $\bm v$ is given by \eqref{eq: vi}.

This Green's function can be interpreted as a transition kernel in a general sense. In terms of an observable  $A$, the action of the Green's function is expressed as follows,
\begin{equation}
  \int A(\bm r') G(\bm r', \bm r,\dt) d\bm r' = A(\bm r)  + \frac{\hbar^2}m \bm v(\bm r) \cdot \nabla A(\bm r) \dt + \frac{\hbar^2}m  \tfrac{\dt}{2} \Laplace A(\bm r) 
  +   \Delta t\big(E_T - {E}_\text{tot}(\bm r)   \big) A(\bm r)+   \mathcal{O}(\dt^2).
\end{equation}

One can write  $G(\bm r', \bm r,\dt) =G_1(\bm r', \bm r,\dt) G_2(\bm r', \bm r,\dt),$ with
\begin{equation}
\begin{aligned}
 G_1(\bm r', \bm r,\dt)  =&  \frac{1}{ \big( 2\pi \sigma^2 \big)^{3N/2}}
     \exp \left[ -  \frac{ \big(\bm r' - \bm r - \frac{h\hbar \Delta t}m \bm v(\bm r)\big)^2 }{2\sigma^2} \right], \\
      G_2(\bm r', \bm r,\dt)  =&  \exp \left[\Delta t \big( E_T - E_\text{tot}(\bm r)  \big) \right].
\end{aligned}
\end{equation}

Computationally, the two operations are carried out in two steps, which can be viewed as an operator-splitting method. Better results are often obtained with a symmetric splitting, which corresponds to redefining,
\begin{equation}\label{eq: G2}
G_2(\bm r', \bm r,\dt)  = \exp \left[ \Delta t\big(E_T -  \tfrac{1}{2}({E}_\text{tot}(\bm r) +{E}_\text{tot}(\bm r'))  \big) \right].
\end{equation}

A typical DMC algorithm begins with an ensemble of $L$ copies of the system, also known as walkers \cite{anderson1975random}. For each realization, one first solves 
the SDEs,
\begin{equation}\label{eq: lgv'}
 d \bm  r_i(t) = \frac{\hbar^2}{{m}}  \nabla \log \phi(\bm r_i) dt + \frac{\hbar^2}{{m}} \sum_{j\ne i} \bm v_{ij} dt + \sigma dW_i(t).
\end{equation}
This step corresponds to the action of the first Green's function $G_1.$ Specifically, $\bm r$ and $\bm r'$ in $G_1$ refer to, respectively, the positions of the particles before and after these SDEs are solved for one time step. 
As alluded to at the beginning of this section, these SDEs coincide with the over-damped Langevin equations \eqref{eq: lgv} after a simple rescaling of the time variable. 

One can think of the approximations by these SDEs as an approximation of the function $f(\bm r, t)$ using a sum of delta functions,
\begin{equation}
  f(\bm r, t) \approx \frac{1}{L} \sum_{\ell=1}^L \delta(\bm r - \bm r^{(\ell)}(t) ).
\end{equation}
   The Green's function $G_1$ is precisely the transition kernel. In particular, the number of walkers will not be changed by this step.

After the particles at the step $t+\dt$ are updated by $G_1$, the Green's function $G_2$ in \eqref{eq: G2} needs to be incorporated. 
This is done by using a birth/death process to determine whether a realization should be removed or duplicated. For each walker, one computes a weight factor,
\begin{equation}
 w(t+\Delta t) = \exp \left[ \Delta t\big(E_T -  \tfrac{1}{2}({E}_\text{tot}(\bm r) +{E}_\text{tot}(\bm r'))  \big) \right], 
\end{equation}
which corresponds to the Green's function $G_2 $ in \eqref{eq: G2}. To apply  Green's function $G_2$, the walkers are duplicated (removed) based on the magnitude of $w(t+\Delta t)$. 
The overall algorithm is summarized on {\bf Algorithm} \ref{alg:dmc}, which will be later referred to as the direct DMC method.

\begin{algorithm}
\caption{Diffusion Monte Carlo (Direct DMC) }
\label{alg:dmc}
\begin{algorithmic}
\STATE{Sample the initial num\_walkers walkers using the  VMC algorithm. Set $M(1)$ as the number of walkers initially. Set $E_T$ to be the average energy computed from the VMC. }  
\smallskip
\FOR{nt=1, num\_steps}
\smallskip
\FOR{n=1, num\_walkers}
\smallskip

    \STATE{Compute the energy ${E}_\text{tot}(\bm r).$}
    
    \STATE{Drift and diffuse the nth walker  according to \eqref{eq: lgv'}.
    %\begin{equation}\label{eq: dd}
    % $   \bm r'\leftarrow \bm r - \nabla V \dt +  \bm v(r) \dt + \sigma \xi.  $
    %\end{equation}
    }
    \STATE{Compute the energy ${E}_\text{tot}(\bm r').$}
        \STATE{Determine the probability of the branching process: $$w_n= \exp \left[ \Delta t \big(E_T- ({E}_\text{tot}(\bm r)+ {E}_\text{tot}(\bm r'))/2 \big)\right].$$  }
    \ENDFOR
    \FOR{$n$=1, num\_walkers}
        %\STATE{ \(w(\bm r) \leftarrow \frac{w(\bm r)}{\sum_{\bm r'} w(\bm r')}\)}
        \IF{ $w_n < 1$ } 
                \STATE{The walker survives with probability $w_n.$}
                \ELSE
                \STATE{The walker is duplicated  $\floor{w_n}$ times. A new walker is created with probability $w_n-\floor{w_n}$.}
                \ENDIF
                \ENDFOR
                \STATE{Recount the number of walkers num\_walkers, and set it to $M(\text{nt}+1)$.}
                
                \STATE{}
                \STATE{Adjust the energy shift: $E_T \leftarrow E_T + \kappa \ln \frac{M(\text{nt}+1)}{M(\text{nt})}. $}
    
\ENDFOR
\end{algorithmic}
\end{algorithm}

\subsection{The random batch algorithm for DMC}
Since the initialization, as well as the drift-diffusion step  of the DMC  involves the solution of the over-damped Langevin dynamics \eqref{eq: lgv} (or \eqref{eq: lgv'}), our random batch algorithm for VMC can be directly applied to this part of the DMC method, to mitigate the same issue encountered in the Metropolis-Hastings algorithm.  

It remains to treat the transition kernel $G_2(\bm r', \bm r, \dt)$ \eqref{eq: G2}. The primary challenge is that computing the energy at each step requires $\mathcal{O}((N+M)N)$ operations in order to update the position of $N$ particles. To reduce this part of the computation,  we propose to write the total energy \eqref{eq: E-tot'} as follows,
 \begin{equation}\label{eq: e-parti}
 {E}_\text{tot}(\bm r)= \sum_{i=1}^N E_1(\bm r_i) + \sum_{1\le i<j\le N} E_2(\bm r_i, \bm r_j) + \sum_{1\le i <j<k\le N} E_3(\bm r_i, \bm r_j, \bm r_k).
\end{equation}
These three terms are onsite, two-body, and three-body contributions.  The on-site energy comes from the one-particle wave function and the external potential,
\begin{equation}\label{eq: E1}
E_1(\bm r_i)= - \frac{\hbar^2}{2m} \nabla^2 \ln \phi(\bm r_i) - \frac{\hbar^2}{2m} |\nabla  \ln \phi(\bm r_i) |^2 +  \sum_{\alpha=1}^{M} U(\bm r_i - R_\alpha).
\end{equation}

To ensure that this part of the energy is evaluated with $\mathcal{O}(1)$ operations,  we pick one atom $\alpha$ in the external potential randomly in the last term, and compute,
\begin{equation}\label{eq: E1'}
E_1(\bm r_i)= - \frac{\hbar^2}{2m} \nabla^2 \ln \phi(\bm r_i) - \frac{\hbar^2}{2m} |\nabla  \ln \phi(\bm r_i) |^2 +  M U(\bm r_i - R_\alpha).
\end{equation}

Let $\bm r_{ij}=\bm r_i -\bm r_j$ be the relative position and $r_{ij}=|\bm r_{ij}|$ be its distance. The two-body term consists of the following terms, 
\begin{equation}\label{eq: E2}
E_2(\bm r_i, \bm r_j)= - \frac{\hbar^2}{m} \nabla^2 \ln u( r_{ij}) + \frac{\hbar^2}{m} \big(\nabla  \ln \phi(\bm r_i)-\nabla  \ln \phi(\bm r_j)\big) \cdot \nabla  u( r_{ij})  + \frac{\hbar^2}{m}  |\nabla  u( r_{ij})|^2 + W( r_{ij}).
\end{equation}

The three-body term can be derived from the first term in the kinetic energy \eqref{eq: Ki}, and it is given by,
\begin{equation}\label{eq: E3}
E_3(\bm r_i, \bm r_j, \bm r_k)=  \frac{\hbar^2}{m} \Big[ \nabla  u( r_{ij}) \cdot \nabla  u( r_{ik}) + \nabla  u(r_{ji}) \cdot \nabla  u( r_{jk}) + \nabla  u( r_{ki}) \cdot \nabla  u( r_{kj}) \Big].
\end{equation}
These three-body terms arise due to the $\|\nabla V\|^2$ term in \eqref{eq: E-tot}.

This partition of the energy is structured in the same manner as in molecular dynamics models \cite{allen2017computer}.  In the random batch algorithm, we randomly pick a batch $C_I$ with three particles:     $C_I=\{ i, j, k\}.$  We first update the position of the three particles (drift and diffuse) by solving the over-damped Langevin dynamics \eqref{eq: lgv'} using the random batch algorithm with batch size 3.  This is demonstrated in \eqref{eq: move-ijk} in {\bf Algorithm \ref{alg:dmc-rb}}. 
We then define a {\it local} energy,
\begin{equation}\label{eq: EIijk}
\begin{aligned}
  {E}_I(\bm r_i, \bm r_j, \bm r_k) =&  E_1(\bm r_i) + E_1(\bm r_j) + E_1(\bm r_k)  \\
          & + \tfrac{N-1}{2} \Big[ E_2(\bm r_i,\bm r_j) + E_2(\bm r_j,\bm r_k) + E_2(\bm r_k,\bm r_i) \Big],\\
          & + \tfrac{(N-1)(N-2)}{2} E_3(\bm r_i,\bm r_j,\bm r_k).   
\end{aligned}
\end{equation} 
In light of \eqref{eq: E1'}, \eqref{eq: E2}, and \eqref{eq: E3}, the cost for evaluating this local energy \eqref{eq: EIijk} remains $\mathcal{O}(1).$

In the branching step of our new DMC method, 
%rather than computing the total energy directly, we randomly pick a batch of three particles. After the particles in the batch are updated,  
we assign a batch with a weight,
\begin{equation}
  w_I = \exp \left[ \dt\big(  \tfrac3N E_T- {E}_I(\bm r_i, \bm r_j, \bm r_k ) \big)\right], 
\end{equation}
which helps to determine whether a walker should be continued/duplicated/deleted. 
This amounts to an approximation of Green's function $G_2$. To see this,
note, on average, the effect of this random procedure on $f(\bm r, t)$ is given by,
\begin{equation}\label{eq: G2-rbm}
\begin{aligned}
  \frac{6}{N(N-1)(N-2)} &\sum_{i<j<k} w_I (\bm r_i, \bm r_j, \bm r_k ) f(\bm r, t) \\
  =& \frac{6}{N(N-1)(N-2)} \sum_{i<j<k}  \Big[ 1  + E_1(\bm r_i)  \dt + E_1(\bm r_j) \dt + E_1(\bm r_k)  \dt \Big]  f(\bm r, t)\\
& +  \frac{3}{N(N-2)} \sum_{i<j<k}  \big( E_2(\bm r_i,\bm r_j) + E_2(\bm r_j,\bm r_k) + E_2(\bm r_k,\bm r_i) \big) \dt  f(\bm r, t)\\
& +  \frac{3}{N} \sum_{i<j<k}  E_3(\bm r_i,\bm r_j,\bm r_k) \dt    f(\bm r, t) + \mathcal{O}(\dt^2),\\
= &  f(\bm r, t) + (E_T - {E}_\text{tot}(\bm r) )  \frac{3\dt}N f +  \mathcal{O}(\dt^2).
\end{aligned}
\end{equation}
Therefore the random batch algorithm is consistent with  Green's function $G_2$ in \eqref{eq: G2} up to order  $\mathcal{O}(\dt^2)$. Note that the evaluation of ${E}_I$ only requires $\mathcal{O}(1)$ operations. 

In the implementation, to avoid frequent removal and duplication of walkers, we apply the branching process after $N/3$ batches of particles are updated. In this case, the weight function is defined by collecting the local energy from each batch (denoted by $I_m$ here),
\begin{equation}
    w(\bm r) = \exp \left[\Delta t \big( E_T  - \widetilde{E}_\text{tot} \big)\right], \quad \widetilde E_\text{tot}=\sum_{m=1}^{N/3} {E}_{I_m}.
\end{equation}
Similar to \eqref{eq: G2-rbm}, one can verify with direct calculations that the branching process with probability $w(\bm r)$ is also consistent with Green's function $G_2$ in \eqref{eq: G2}.
Overall, the algorithm is summarized in {\bf Algorithm 4}.

\begin{algorithm}
\caption{Diffusion Monte Carlo using Random Batch (RBM-DMC) }
\label{alg:dmc-rb}
\begin{algorithmic}
\STATE{Sample the initial num\_walkers walkers using a VMC algorithm. Set $M(1)$ to be the number of walkers initially. Set $E_T$ to be the average energy computed from the VMC.}  
\smallskip
\FOR{nt=1, num\_steps}
\smallskip
\FOR{n=1, num\_walkers}
\smallskip

\FOR{m=1, N/3}
    \STATE{Randomly pick a batch $I_m$ with three particles $(i, j, k)$.}
       \STATE{Perform one step of the Monte Carlo algorithm with respect to  $\left\{q_\alpha^i\right\}$ and select $\alpha$. Compute $\bm b_i=- \nabla \theta(\bm r_i - R_\alpha)$. Similarly compute $\bm b_j$ and $\bm b_k$.}  
     \STATE{Evaluate $\bm u_{ij}= -\bm u_{ji}=  (N-1) \nabla_{\bm r_i} u(|\bm r_i - \bm r_j|).$ Similarly evaluate $\bm u_{ik}$ and $\bm u_{jk}$.}
    \STATE{Update the position of the three particles,
    \begin{equation}\label{eq: move-ijk}
 \begin{aligned}
  {\bm r}_i \longleftarrow &{\bm r}_i +  \frac{\hbar^2}m \bm b_i \dt
    + \frac{\hbar^2}m (\bm u_{ij} +  \bm u_{ik}) \dt + \sigma  \Delta W_i, \\
   {\bm r}_j \longleftarrow &{\bm r}_j + \frac{\hbar^2}m \bm b_j \dt  + \frac{\hbar^2}m (\bm u_{ji} + \bm u_{jk}) \dt + \sigma  \Delta W_j, \\
   {\bm r}_k \longleftarrow &{\bm r}_k + \frac{\hbar^2}m \bm b_k \dt  + \frac{\hbar^2}m (\bm u_{ki} + \bm u_{kj}) \dt +  \sigma \Delta W_k. \\
    \end{aligned}
\end{equation}
    \STATE{Compute the local batch energy ${E}_{I_m}\big(\bm r(t+\dt)\big)$ from \eqref{eq: EIijk}. } 
}
\ENDFOR
    
        \STATE{Determine the probability of the branching process from $E_n,$ $E_n=\sum_{m=1}^{N/3} {E}_{I_m},$ $$w_n= \exp \left[\Delta t \big( E_T  - {E}_n \big)\right].$$  }
    \ENDFOR

   \STATE{Branch the walkers and adjust the energy $E_T$ as in the direct DMC algorithm}
\ENDFOR

\end{algorithmic}
\end{algorithm}

\subsection{Numerical Results}

Now we test the RBM-DMC ({\bf Algorithm} \ref{alg:dmc-rb}) and compare the results with the direct DMC method ({\bf  Algorithm} \ref{alg:dmc}). For the initialization, we first apply a VMC method using the ansatz \eqref{eq: ansatz} for the wave function $\Phi_0.$ The Metropolis-Hastings Monte Carlo method is used in both methods so that they start at the same states. 300 ensembles are created by sub-sampling one sample out of every 500 steps from the VMC runs to avoid correlations among the ensembles. For both methods, we use $\dt=10^{-4}$ and run $200,000$ steps of  simulations.   
%This initialization step can also be done by using the RBM method.  

\begin{figure}[hptb]
\centering
    \includegraphics[scale=0.1]{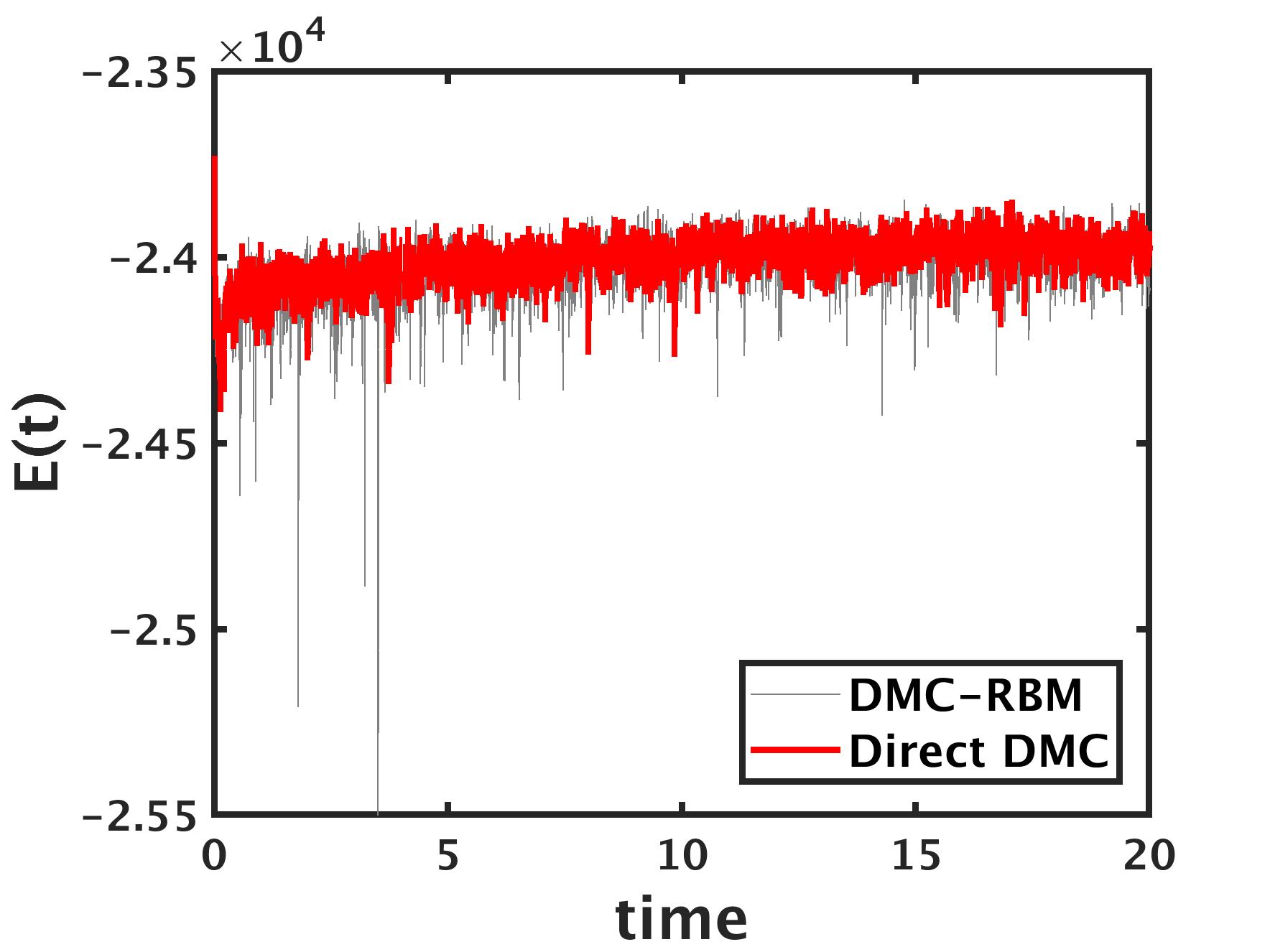}\\
        \includegraphics[scale=0.1]{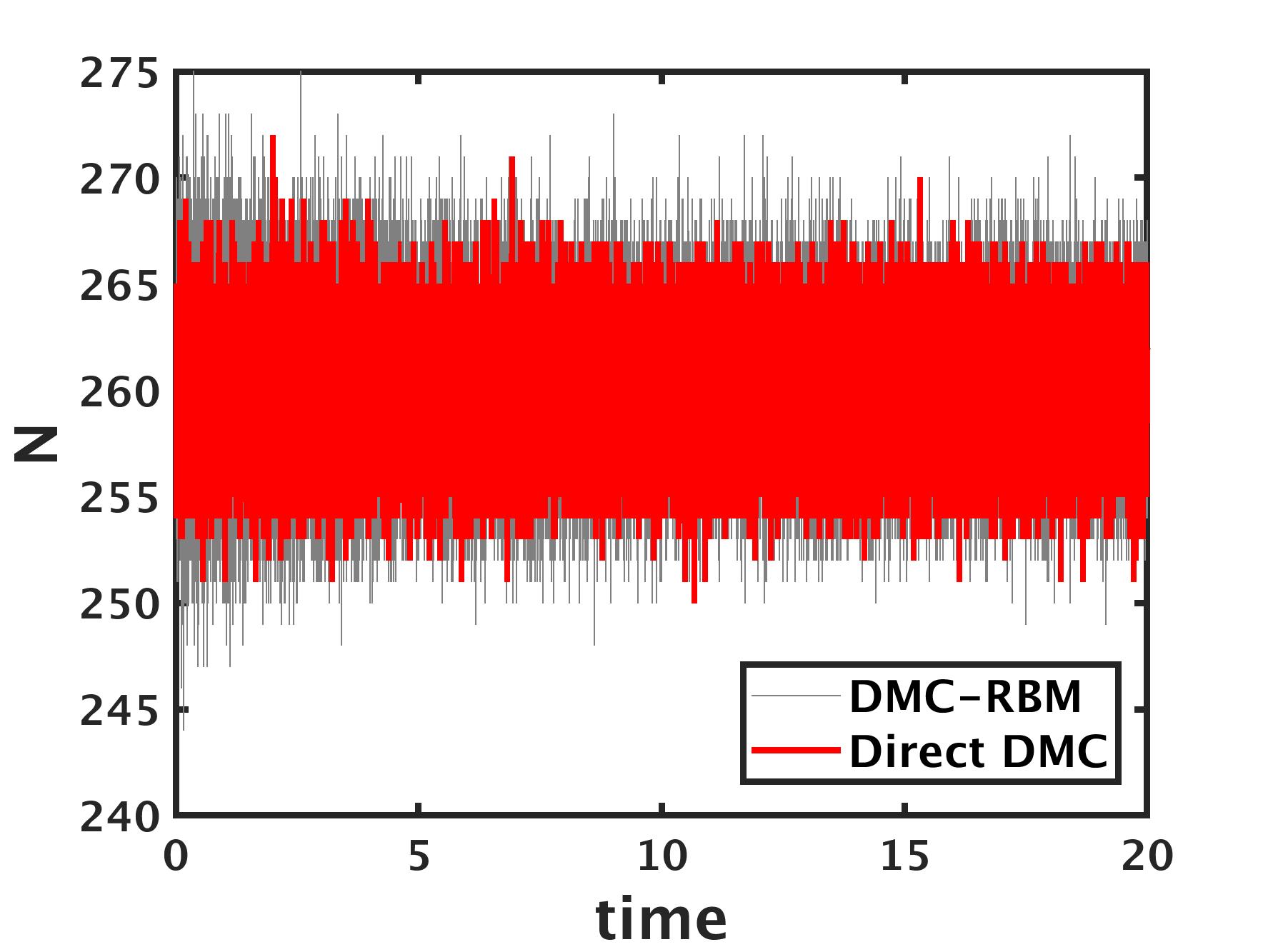}\\
            \includegraphics[scale=0.1]{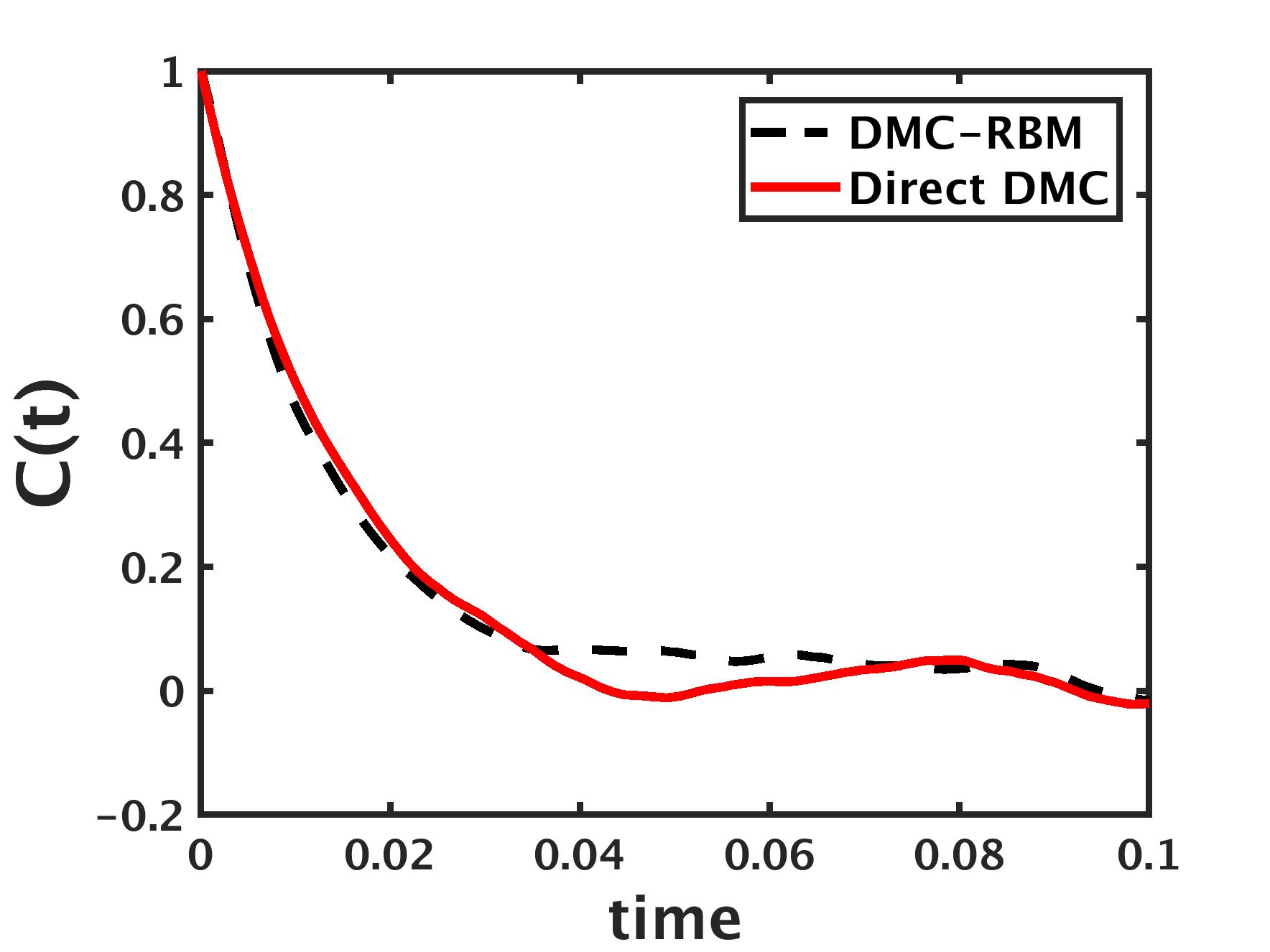}
    \caption{A comparison of the RBM-DMC ({\bf  Algorithm} \ref{alg:dmc-rb}) to the direct DMC  method ({\bf Algorithm} \ref{alg:dmc}). Top: time series; Middle: The number of walkers; Bottom: time correlation. }
    \label{fig:dmc}
\end{figure}

 Figure \ref{fig:dmc} shows the time series (top panel) generated by the two algorithms. We observe that the random batch method generates samples with slightly larger fluctuations during the burn in period. But the fluctuations eventually become comparable to those from the direct DMC simulations. The population of the walkers (middle panel) exhibits a similar behavior.  We also examined the time correlation of the total energy \eqref{eq: E-tot'}. This is done by using the time series within the time interval $(10,20)$ and regard it as a stationary process.

\begin{figure}[htbp]
\centering
    \includegraphics[scale=0.18]{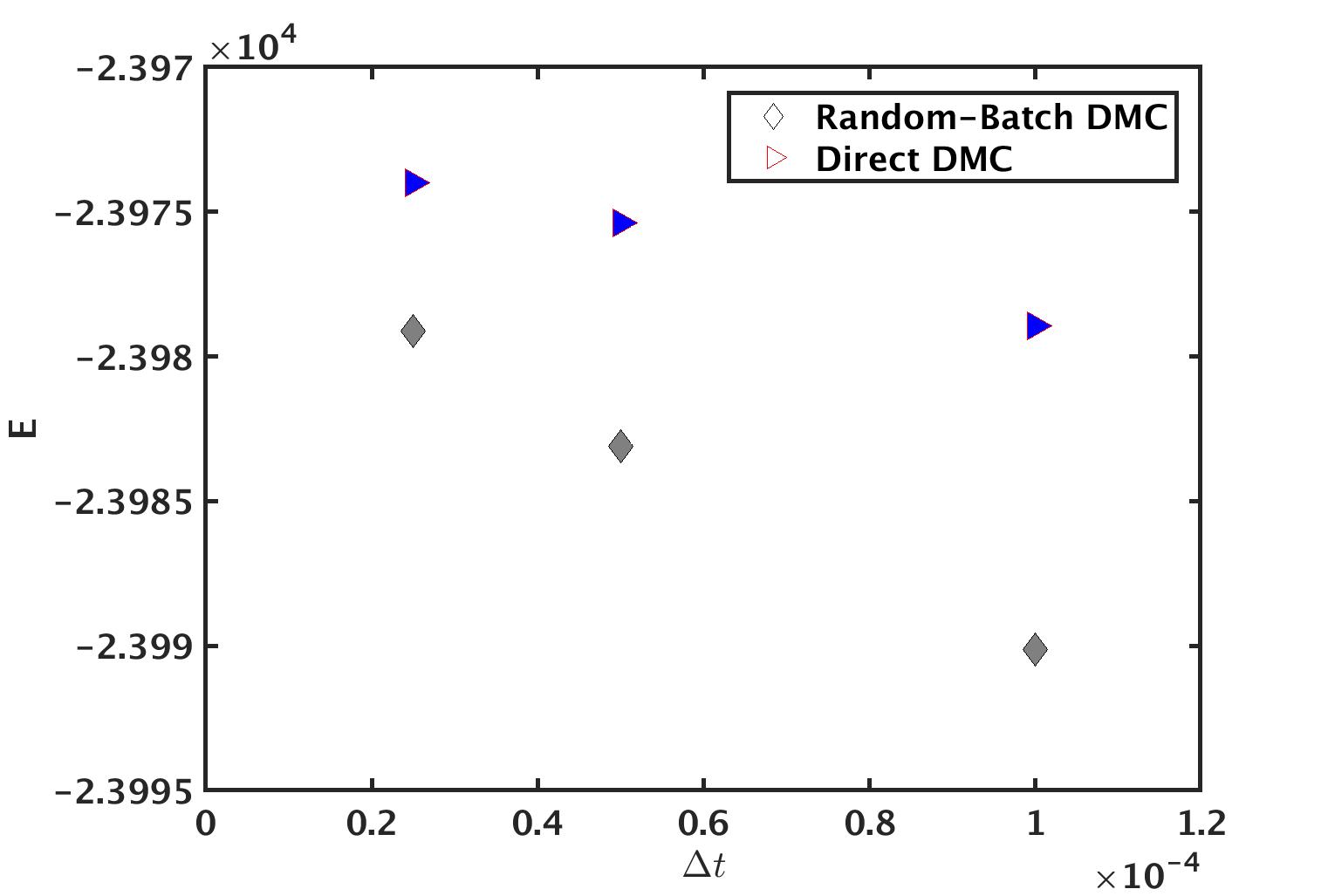}
    \caption{The computed average energy for several choices of the step size $\dt$ . }
    \label{fig:dmc_err}
\end{figure}

We conduct simulations with various choices of the step size $\Delta t$ to monitor the convergence. Figure \ref{fig:dmc_err} shows the energy computed from each instance. We decreased $\Delta t$ from $ 10^{-4}$ to $0.5\times10^{-4}$, and then further to 
 $0.25\times10^{-4}.$ We observe that the results from the direct DMC and the random batch DMC methods both exhibit linear convergences. The extrapolated energy values at $\dt =0$ are $-2.39723\times 10^4$ and $-2.39756\times 10^4$, respectively. 

\begin{figure}[htbp]
\centering
    \includegraphics[scale=0.09]{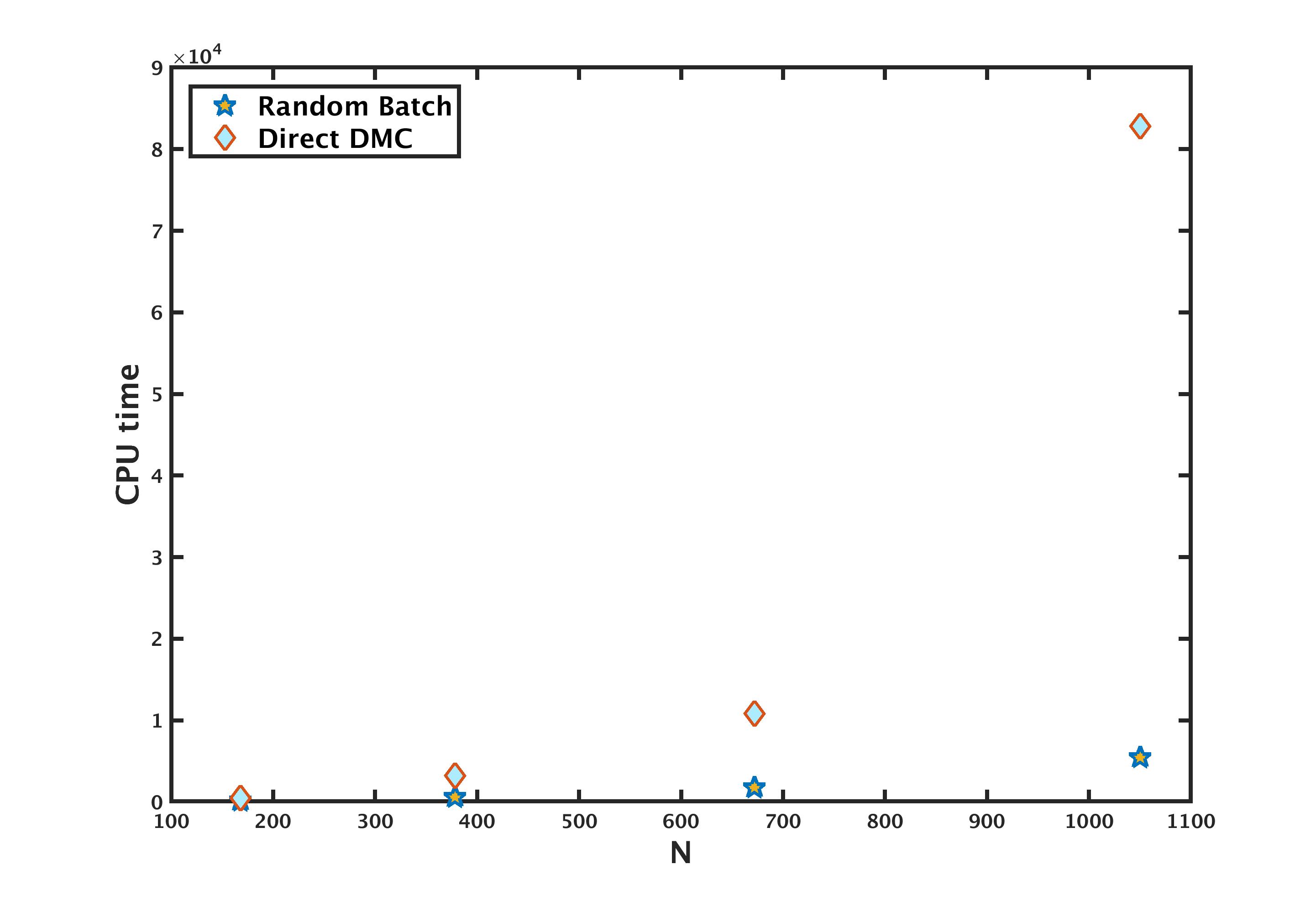}
    \caption{A comparison of the CPU runtime (in seconds) for running 1000 steps of DMC. }
    \label{fig:CPU}
\end{figure}

Since our primary focus is on the speedup of the computation, We  examine the CPU runtime for various system sizes. More specifically, we increase the system size from the original 168 particles, to $N= 378$,  $N=672$ and $N=1050$ particles, and in each case, we run the direct DMC and the RBM-DMC for 1000 steps. For the initial system $N=168$, the runtimes are 129.29 and 474.44 (seconds) for RBM-DMC and direct DMC, respectively. In this case, the random batch algorithm requires 1/4 of the CPU time, which is a moderate speedup. But  as shown in Figure \ref{fig:CPU}, the CPU time for the direct DMC method increases much more rapidly as $N$ increases. 

\begin{figure}[htbp]
\centering
    \includegraphics[scale=0.3]{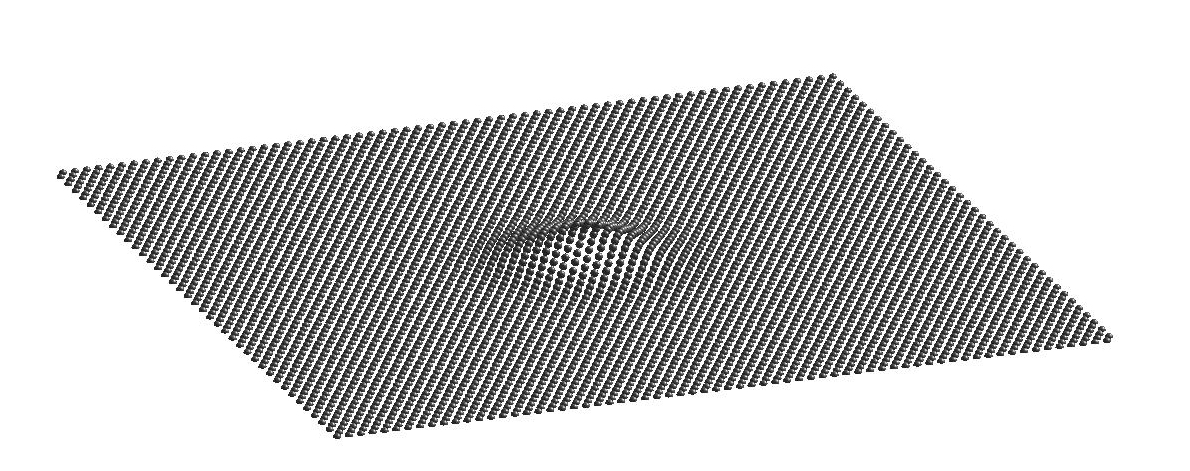}
    \caption{The out-of-plane displacement of the atoms on the graphite lattice. }
    \label{fig:dsp}
\end{figure}

With the advent of modern high-performance computer clusters, QMC methods have become a leading candidate for computing electronic structures of relatively large systems. As demonstrated in \cite{kim2018qmcpack}, direct DMC methods can be implemented in multi-core processors, by distributing the random walkers among different units. As a first step toward this goal, we study the $^4$He system on a graphite lattice with non-homogeneous deformation. More specifically, by mimicking an external load, we displace the atoms in the third direction according to a Gaussian profile:
\begin{equation}\label{eq: defm}
 z_j = z_e +  h_0 \exp \left[ -(x_j^2 + y_j^2)/1000\right],      
\end{equation}
with $h_0$ indicating the height of the sheet at the origin. To establish such a spatial profile, a much larger system is needed. We consider a system with 5016 atoms, as shown in Figure \ref{fig:dsp}. We implemented RBM-DMC ({\bf Algorithm}  \ref{alg:dmc-rb}) on 60 CPUs by distributing the walkers among the CPUs. After each branching step, the walkers are re-distributed to maintain a load balance. 

\begin{figure}[htbp]
\centering
    \includegraphics[scale=0.23]{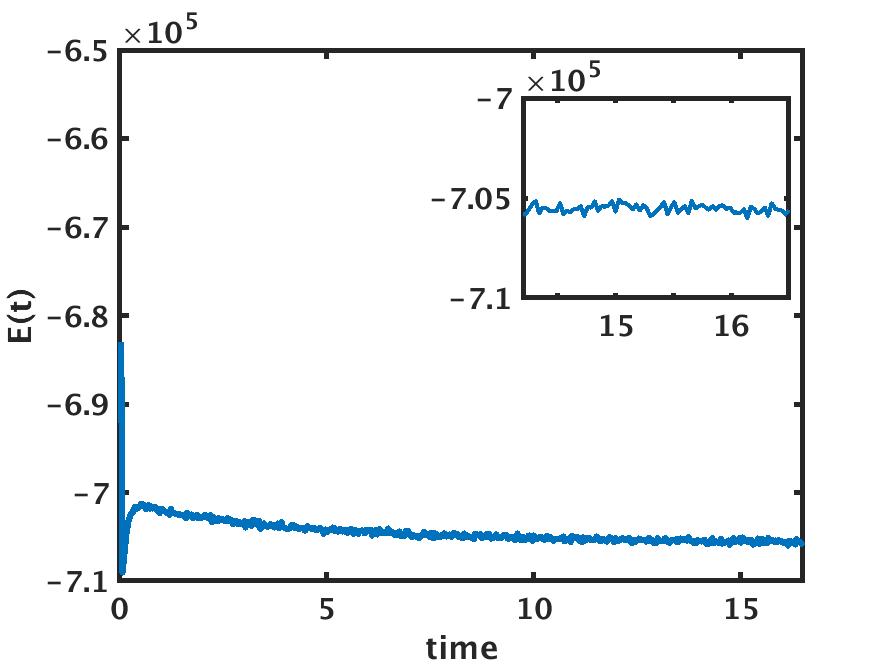}
        \includegraphics[scale=0.23]{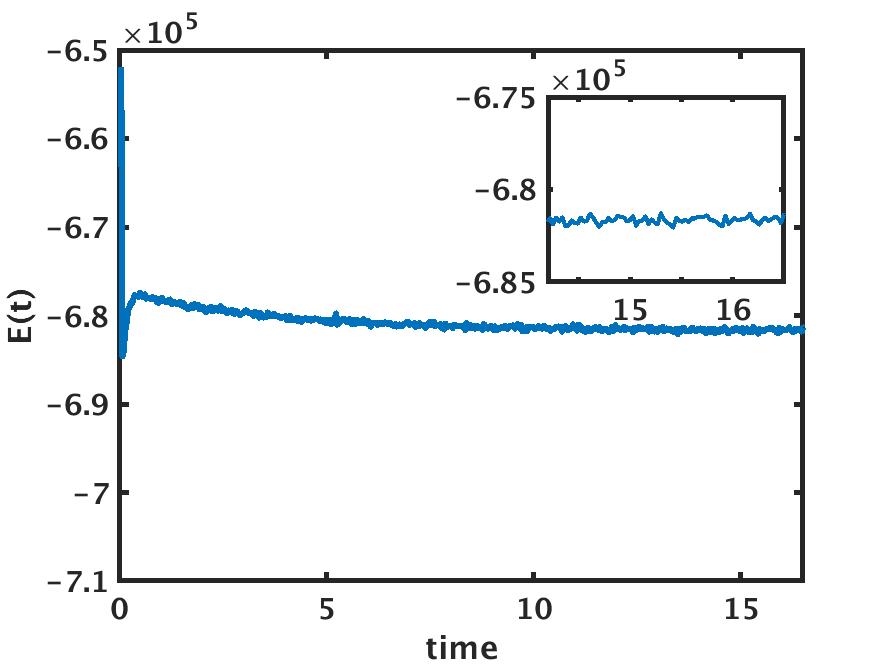}
    \caption{The energy from the VMC simulations. Left: undeformed lattice; Right: with deformation \eqref{eq: defm}. The insets show the energy after the system reaches equilibrium.  }
    \label{fig:vmcL}
\end{figure}

We first perform the VMC simulations with 180 ensembles on the two systems, including the homogeneous lattice ($h_0=0),$ and the deformed lattice (we pick $h_0=2a_0$). This is done by using the RBM-DMC ({\bf Algorithm} \ref{alg:dmc-rb}) with the branching process turned off.  We choose $\Delta t=10^{-4} $ and run the algorithms for $160,000$ steps. Figure \ref{fig:vmcL} shows the energy computed from the iterations and averaged over the 180 ensembles. In both cases, the energy exhibits a sharp relaxation before reaching a steady profile. We notice that the deformation leads to higher ground state energy. Each of the VMCs simulations take about 30 hours.

\begin{figure}[htbp]
\centering
    \includegraphics[scale=0.156]{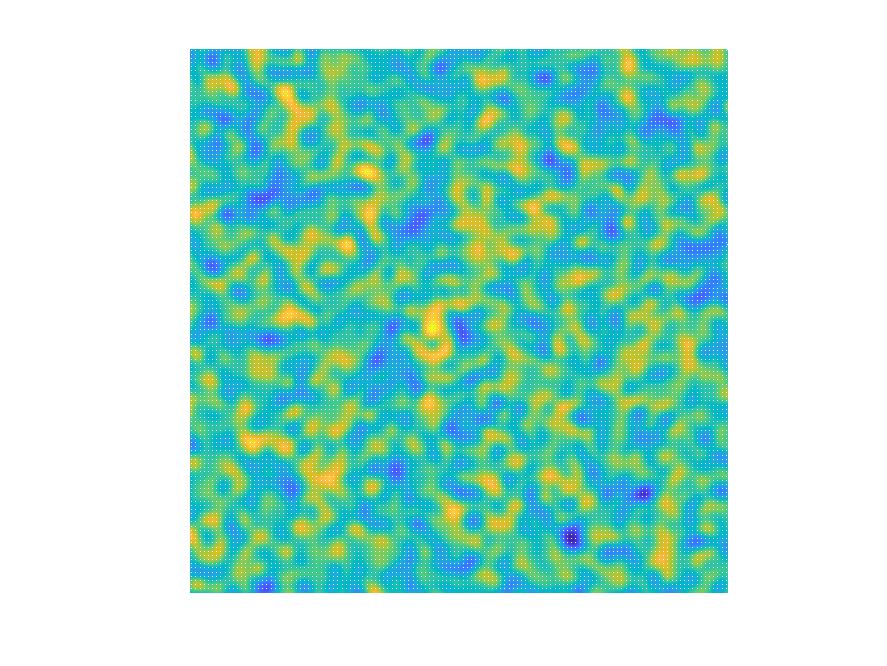}
        \includegraphics[scale=0.156]{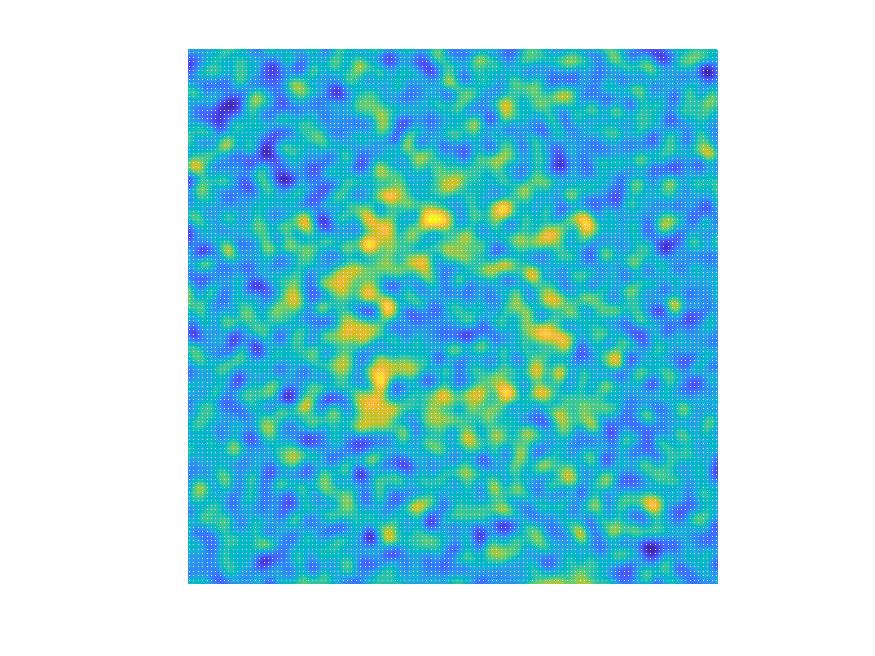}
        \includegraphics[scale=0.156]{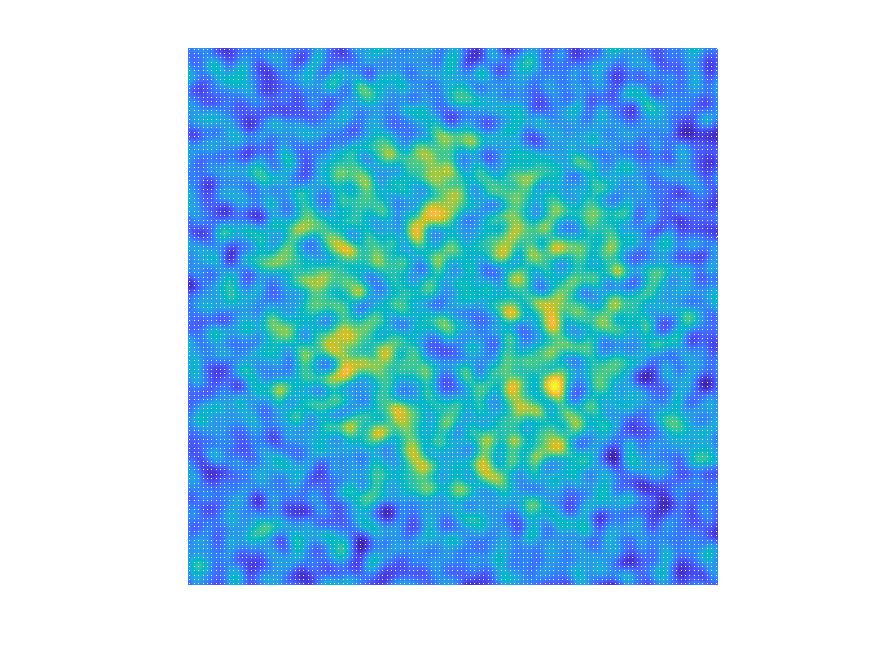}
    \caption{The particle density. Left: undeformed lattice after the VMC sampling; Middle: System with deformation after the VMC sampling; Right: System with deformation after the DMC sampling. }
    \label{fig:den}
\end{figure}

At the end of the VMC run, we computed the particle density, from the 180 ensembles. For visualization purpose, we use the smoothed-kernel density estimator (mvksdensity in MATLAB) with width $1.5$\AA~  to obtain the density. In this method, the position of each particle (out of 5016) is interpreted as a data point, and the kernel density includes the contribution from all particles and all the ensembles.  Figure \ref{fig:den} shows the density plots for both cases. An interesting observation is that in the deformed case, higher density is found in an annulus region, where the deformation is the largest.

\begin{figure}[htbp]
\centering
    \includegraphics[scale=0.34]{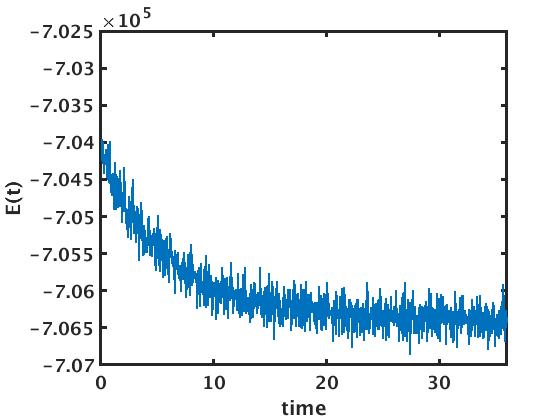}
        \includegraphics[scale=0.34]{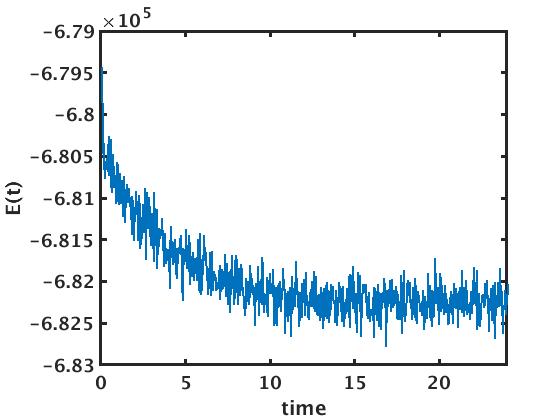}
    \caption{The energy from the DMC simulations. Left: undeformed lattice; Right: with deformation. }
    \label{fig:dmcL}
\end{figure}

With the walkers prepared by the VMC simulation, we perform DMC simulations with the RBM-DMC method ({\bf algorithm} \ref{alg:dmc-rb}). Again we use $\dt=10^{-4}$ and we ran 240,000 steps of the algorithm. We monitor the energy and Figure \ref{fig:dmcL} shows  how the energy changes during the simulations.  The system with homogeneous lattice takes slightly longer to reach the steady state, and therefore we run the simulation for an extended period (360,000 steps).  
 
\section{Summary and Discussions}
We have constructed random batch algorithms for quantum Monte Carlo simulations. The main objective is to alleviate the computational cost associated with the calculations of two-body interactions, including the particle interactions in the potential energy,  and the pairwise terms in the Jastrow factor. In the framework of variational Monte Carlo methods, the random batch algorithm is constructed based on the over-damped Langevin dynamics, so that updating the position of each particle only requires $\mathcal{O}(1)$ operations per time step. Consequently for the N-particle system the computational cost per time step is reduced from $O(N^2)$ to $O(N)$.
For the diffusion Monte Carlo method, we proposed to decompose the total energy into on-site, two-body, and three-body terms, which can be evaluated within a random batch of three particles. This still guarantees   $\mathcal{O}(N)$ operations per time step for the $N$-body  particle system. 

We have placed the main emphasis on the speedup of the computation. The speedup is more significant for larger systems, where the asymptotic scaling kicks in.  In terms of the accuracy, we have shown that the random algorithms have first-order accuracy, comparable to the Euler-Maruyama method. This is certainly a low-order method. For instance, in the VMC simulations, we observed that the random batch algorithm remains stable when $\dt=0.05,$ but the step size has to be reduced to at least $\dt = 0.001$ to ensure a good accuracy. In this case, high-order diffusion Monte Carlo methods \cite{forbert2001fourth} would be helpful, and the construction of random batch algorithms with higher accuracy is certainly an open issue. Another common practice to correct the bias is to combine the algorithm with an Metropolis-Hastings step to accept/reject samples generated by the random batch method \cite{reynolds1982fixed,scemama2006efficient}. Maintaining detailed balance in the random batch algorithm is another interesting direction.  

In principle, some of these interactions in QMC can be (and have been) treated using fast summation methods, \eg, the fast multipole methods for Coulomb interactions or Gaussian functions \cite{cheng1999fast,greengard1991fast}. But compared to the fast summation methods, the implementation of RBM is much easier. 

This paper only focuses on the VMC and DMC methods. 
Another important methodology is the path-integral quantum Monte Carlo \cite{herman1982path,sarsa2000path,ceperley1995path}, which works with the density-matrix at finite temperature. The formulation of path integral method using molecular dynamics techniques \cite{tuckerman1993efficient} seems to be an appropriate platform to implement the RBM.

\section*{Acknowledgment}

Jin's research is partly supported by NSFC grant No. 11871297. Li's research is supported by NSF under grant DMS-1819011 and DMS-1953120. 

\bibliographystyle{plain}
\bibliography{qmc}

\begin{thebibliography}{10}

\bibitem{allen2017computer}
Michael~P Allen and Dominic~J Tildesley.
\newblock {\em Computer Simulation of Liquids}.
\newblock Oxford university press, 2017.

\bibitem{anderson1975random}
James~B Anderson.
\newblock A random-walk simulation of the {Schr\"odinger} equation: {H}+3.
\newblock {\em The Journal of Chemical Physics}, 63(4):1499--1503, 1975.

\bibitem{anderson2007quantum}
James~B Anderson.
\newblock {\em Quantum {Monte Carlo:} origins, development, applications}.
\newblock Oxford University Press, 2007.

\bibitem{bottou1998online}
L{\'e}on Bottou.
\newblock Online learning and stochastic approximations.
\newblock {\em On-line learning in neural networks}, 17(9):142, 1998.

\bibitem{bubeck2014convex}
S{\'e}bastien Bubeck.
\newblock Convex optimization: Algorithms and complexity.
\newblock {\em arXiv preprint arXiv:1405.4980}, 2014.

\bibitem{carleo2017solving}
Giuseppe Carleo and Matthias Troyer.
\newblock Solving the quantum many-body problem with artificial neural
  networks.
\newblock {\em Science}, 355(6325):602--606, 2017.

\bibitem{ceperley1995path}
David~M Ceperley.
\newblock Path integrals in the theory of condensed helium.
\newblock {\em Reviews of Modern Physics}, 67(2):279, 1995.

\bibitem{cheng1999fast}
Hongwei Cheng, Leslie Greengard, and Vladimir Rokhlin.
\newblock A fast adaptive multipole algorithm in three dimensions.
\newblock {\em Journal of computational physics}, 155(2):468--498, 1999.

\bibitem{weinan2005analysis}
Weinan E, Di~Liu, and Eric Vanden-Eijnden.
\newblock Analysis of multiscale methods for stochastic differential equations.
\newblock {\em Communications on Pure and Applied Mathematics},
  58(11):1544--1585, 2005.

\bibitem{forbert2001fourth}
Harald~A Forbert and Siu~A Chin.
\newblock Fourth-order diffusion {Monte Carlo} algorithms for solving quantum
  many-body problems.
\newblock {\em Physical Review B}, 63(14):144518, 2001.

\bibitem{foulkes2001quantum}
WMC Foulkes, Lubos Mitas, RJ~Needs, and G~Rajagopal.
\newblock Quantum {Monte Carlo} simulations of solids.
\newblock {\em Reviews of Modern Physics}, 73(1):33, 2001.

\bibitem{frenkel2001understanding}
Daan Frenkel and Berend Smit.
\newblock {\em Understanding molecular simulation: from algorithms to
  applications}, volume~1.
\newblock Elsevier, 2001.

\bibitem{GJP}
Francois Golse, Shi Jin, and Thierry Paul.
\newblock The random batch method for $n$-body quantum dynamics.
\newblock {\em arXiv:1912.07424}, 2020.

\bibitem{greengard1991fast}
Leslie Greengard and John Strain.
\newblock The fast {Gauss} transform.
\newblock {\em SIAM Journal on Scientific and Statistical Computing},
  12(1):79--94, 1991.

\bibitem{han2020solving}
Jiequn Han, Jianfeng Lu, and Mo~Zhou.
\newblock Solving high-dimensional eigenvalue problems using deep neural
  networks: A diffusion {Monte Carlo} like approach.
\newblock {\em arXiv preprint arXiv:2002.02600}, 2020.

\bibitem{han2019solving}
Jiequn Han, Linfeng Zhang, and E~Weinan.
\newblock Solving many-electron {Schr\"odinger} equation using deep neural
  networks.
\newblock {\em Journal of Computational Physics}, 399:108929, 2019.

\bibitem{herman1982path}
MF~Herman, EJ~Bruskin, and BJ~Berne.
\newblock On path integral {Monte Carlo} simulations.
\newblock {\em The Journal of Chemical Physics}, 76(10):5150--5155, 1982.

\bibitem{jastrow1955many}
Robert Jastrow.
\newblock Many-body problem with strong forces.
\newblock {\em Physical Review}, 98(5):1479, 1955.

\bibitem{jin2020mean}
Shi Jin and Lei Li.
\newblock On the mean field limit of random batch method for interacting
  particle systems.
\newblock {\em arXiv preprint arXiv:2005.11740}, 2020.

\bibitem{JLL2}
Shi Jin, Lei Li, and Jian-Guo Liu.
\newblock Convergence of random batch method for interacting particles with
  disparate species and weights.
\newblock {\em arXiv:2003.13064}, 2020.

\bibitem{jin2020random}
Shi Jin, Lei Li, and Jian-Guo Liu.
\newblock Random batch methods {(RBM)} for interacting particle systems.
\newblock {\em Journal of Computational Physics}, 400:108877, 2020.

\bibitem{joly1992helium}
F~Joly, C~Lhuillier, and B~Brami.
\newblock The helium-graphite interaction.
\newblock {\em Surface science}, 264(3):419--422, 1992.

\bibitem{kalos1974helium}
Malvin~H Kalos, Dominique Levesque, and Loup Verlet.
\newblock Helium at zero temperature with hard-sphere and other forces.
\newblock {\em Physical Review A}, 9(5):2178, 1974.

\bibitem{kim2018qmcpack}
Jeongnim Kim, Andrew~D Baczewski, Todd~D Beaudet, Anouar Benali, M~Chandler
  Bennett, Mark~A Berrill, Nick~S Blunt, Edgar Josu{\'e}~Landinez Borda,
  Michele Casula, David~M Ceperley, et~al.
\newblock {QMCPACK}: an open source ab initio quantum monte carlo package for
  the electronic structure of atoms, molecules and solids.
\newblock {\em Journal of Physics: Condensed Matter}, 30(19):195901, 2018.

\bibitem{kloeden2013numerical}
Peter~E Kloeden and Eckhard Platen.
\newblock {\em Numerical solution of stochastic differential equations},
  volume~23.
\newblock Springer Science \& Business Media, 2013.

\bibitem{KZ2020}
Dongnam Ko and Enrique Zuazua.
\newblock Model predictive control with random batch methods for a guiding
  problem.
\newblock {\em arXiv:2004.14834}, 2020.

\bibitem{Kohn1965}
W.~Kohn and L.~J. Sham.
\newblock {Self-consistent equations including exchange and correlation
  effects}.
\newblock {\em Physical Review}, 140(4A):A1133--A1138, 1965.

\bibitem{li2020stochastic}
Lei Li, Yingzhou Li, Jian-Guo Liu, Zibu Liu, and Jianfeng Lu.
\newblock A stochastic version of stein variational gradient descent for
  efficient sampling.
\newblock {\em Communications in Applied Mathematics and Computational
  Science}, 15(1):37--63, 2020.

\bibitem{LLT}
Lei Li, Jian-Guo Liu, and Yijia Tang.
\newblock A direct simulation approach for the {Poisson-Boltzmann} equation
  using the random batch method.
\newblock {\em arXiv:2004.05614}, 2020.

\bibitem{li2020random}
Lei Li, Zhenli Xu, and Yue Zhao.
\newblock A random-batch {Monte Carlo} method for many-body systems with
  singular kernels.
\newblock {\em SIAM Journal on Scientific Computing}, 42(3):A1486--A1509, 2020.

\bibitem{Mattingly:02}
Jonathan~C Mattingly, Andrew~M Stuart, and Desmond~J Higham.
\newblock Ergodicity for {SDEs} and approximations: locally {Lipschitz} vector
  fields and degenerate noise.
\newblock {\em Stochastic processes and their applications}, 101(2):185--232,
  2002.

\bibitem{mcmillan1965ground}
William~Lauchlin McMillan.
\newblock Ground state of liquid he4.
\newblock {\em Physical Review}, 138(2A):A442, 1965.

\bibitem{needs2020variational}
RJ~Needs, MD~Towler, ND~Drummond, Pablo Lopez~Rios, and JR~Trail.
\newblock Variational and diffusion quantum {Monte Carlo} calculations with the
  casino code.
\newblock {\em The Journal of Chemical Physics}, 152(15):154106, 2020.

\bibitem{pang2014diffusion}
Tao Pang.
\newblock Diffusion {Monte Carlo:} a powerful tool for studying quantum
  many-body systems.
\newblock {\em American Journal of Physics}, 82(10):980--988, 2014.

\bibitem{pfau2019ab}
David Pfau, James~S Spencer, Alexander G de~G Matthews, and W~Matthew~C
  Foulkes.
\newblock Ab-initio solution of the many-electron {Schr\"odinger} equation with
  deep neural networks.
\newblock {\em arXiv preprint arXiv:1909.02487}, 2019.

\bibitem{raftery1992practical}
Adrian~E Raftery and Steven~M Lewis.
\newblock Practical {Markov Chain Monte Carlo}: one long run with diagnostics:
  implementation strategies for {Markov Chain Monte Carlo}.
\newblock {\em Statistical science}, 7(4):493--497, 1992.

\bibitem{reynolds1982fixed}
Peter~J Reynolds, David~M Ceperley, Berni~J Alder, and William~A Lester~Jr.
\newblock Fixed-node quantum {Monte Carlo} for molecules.
\newblock {\em The Journal of Chemical Physics}, 77(11):5593--5603, 1982.

\bibitem{sarsa2000path}
A~Sarsa, KE~Schmidt, and WR~Magro.
\newblock A path integral ground state method.
\newblock {\em The Journal of Chemical Physics}, 113(4):1366--1371, 2000.

\bibitem{scemama2012qmc}
Anthony Scemama, Michel Caffarel, Emmanuel Oseret, and William Jalby.
\newblock Qmc= chem: {A quantum monte carlo program} for large-scale
  simulations in chemistry at the petascale level and beyond.
\newblock In {\em International Conference on High Performance Computing for
  Computational Science}, pages 118--127. Springer, 2012.

\bibitem{scemama2006efficient}
Anthony Scemama, Tony Leli{\`e}vre, Gabriel Stoltz, Eric Canc{\`e}s, and Michel
  Caffarel.
\newblock An efficient sampling algorithm for variational {Monte Carlo}.
\newblock {\em The Journal of chemical physics}, 125(11):114105, 2006.

\bibitem{tuckerman1993efficient}
Mark~E Tuckerman, Bruce~J Berne, Glenn~J Martyna, and Michael~L Klein.
\newblock Efficient molecular dynamics and hybrid {Monte Carlo} algorithms for
  path integrals.
\newblock {\em The Journal of Chemical Physics}, 99(4):2796--2808, 1993.

\bibitem{von1992quantum}
Wolfgang von~der Linden.
\newblock A quantum {Monte Carlo} approach to many-body physics.
\newblock {\em Physics Reports}, 220(2-3):53--162, 1992.

\bibitem{whitlock1998monte}
PA~Whitlock, GV~Chester, and B~Krishnamachari.
\newblock Monte carlo simulation of a helium film on graphite.
\newblock {\em Physical Review B}, 58(13):8704, 1998.

\bibitem{wright2015coordinate}
Stephen~J Wright.
\newblock Coordinate descent algorithms.
\newblock {\em Mathematical Programming}, 151(1):3--34, 2015.

\end{thebibliography}

\end{document}